\documentclass[useAMS,usenatbib]{mn2e}

\usepackage[english]{babel}
\usepackage[dvips]{graphicx}

\title[S0 galaxies in Fornax: data and kinematics]
{S0 galaxies in Fornax: data and kinematics}
\author[A.G. Bedregal, A. Arag\'on-Salamanca, M.R. Merrifield and B. Milvang-Jensen]
{A.G. Bedregal$^{1}$\thanks{E-mail:ppxapgg@nottingham.ac.uk}, 
A. Arag\'on-Salamanca$^{1}$, M.R. Merrifield$^{1}$ and B. Milvang-Jensen$^{2}$ \\
$^{1}$School of Physics and Astronomy, Centre of Astronomy and Particle
Theory, University of Nottingham, \\University Park, Nottingham, NG7 2RD, UK
\\
$^{2}$Dark Cosmology Centre, Niels Bohr Institute, University of Copenhagen, Juliane Maries Vej 30, DK-2100 Copenhagen, Denmark}
\begin{document}

\date{Accepted ***. Received ***; in original form ***}

\pagerange{\pageref{firstpage}--\pageref{lastpage}} \pubyear{2006}

\maketitle

\label{firstpage}

\begin{abstract}

We have obtained long-slit spectroscopy for a sample of 9 S0 galaxies
in the Fornax Cluster using the FORS2 spectrograph at the 8.2m ESO
VLT.  From these data, we have extracted the kinematic parameters,
comprising the mean velocity, velocity dispersion and higher-moment
$h_3$ and $h_4$ coefficients, as a function of position along the
major axes of these galaxies.  Comparison with published kinematics
indicates that earlier data are often limited by their lower
signal-to-noise ratio and relatively poor spectral resolution.  The
greater depth and higher dispersion of the new data mean that we reach
well beyond the bulges of these systems, probing their disk kinematics
in some detail for the first time.  Qualitative inspection of the
results for individual galaxies shows that they are not entirely
simple systems, perhaps indicating a turbulent past.  Nonetheless, we
are able to derive reliable circular velocities for most of these
systems, which points the way toward a study of their Tully--Fisher
relation.  This study, along with an analysis of the stellar
populations of these systems out to large galactocentric distances,
will form the bases of future papers exploiting these new high-quality
data, hopefully shedding new light on the evolutionary history of
these systems.

\end{abstract}


\section{Introduction}
One of the key areas of research in extragalactic astronomy is the
study of the formation and evolution of galaxies in different
environments, from the low-density field to rich clusters.  Our
understanding of the individual physical processes involved in the
evolutionary history of galaxies, and their relative importance in
each environment, is still rather poor.  In this context, lenticular
(S0) galaxies have a very important role.  Located between elliptical
and spiral galaxies in the Hubble Diagram, S0 galaxies have been a
focus of debate for many years. One fundamental -- and sometimes
controversial -- issue is whether the formation of these galaxies is
more closely linked to that of ellipticals or to that of spirals.

The evolution with redshift of the morphology--density relation in
clusters of galaxies (Dressler 1980; Dressler et al.\ 1997) implies
that the relative fraction of S0 galaxies increased from $z\approx
0.5$ to the present, while the relative fraction of spirals decreased
in a similar proportion. An evolutionary connection between S0s and
spirals has thus been proposed, at least in cluster environments. More
recent works seem to support such ideas (e.g., Fasano et al.\ 2000;
Desai et al.\ 2006), although alternative views do exist (Andreon
1998).  However, our understanding of the mechanisms involved in the
possible morphological transformation of spirals into S0s remains
poor.  In addition, close similarities exist between ellipticals and
S0s in their colours, stellar populations, gas content, location on
the fundamental plane, etc, so the debate as to whether S0s are more
closely related to spirals or ellipticals remains open.

The presence of S0s in both cluster and field environments raises the
real possibility that multiple evolutionary paths exist for the
formation of these systems.  Indeed, a variety of mechanisms have been
proposed that can, in principle, alter a galaxy's morphology in this
direction.  These mechanisms include ram pressure gas stripping (Gunn
\& Gott 1972), harassment (Moore et al. 1998), starvation (Larson, Tinsley \&
Caldwell 1980; Bekki et
al. 2002), unequal mass galaxy mergers (Mihos \& Hernquist 1994; Bekki
1998) and galaxy-cluster interactions (Goto et al. 2003).  However, it
is still unclear whether any of these processes would work in
practice, and, if more than one, their relative importance in
different environments.

Van den Bergh (1990), analysing the Revised Shapley--Ames Catalog of
Bright Galaxies (Sandage \& Tammann 1987), found that the frequency
distribution of the luminosity of S0s is not intermediate between E
and Sa galaxies. This discontinuity could imply the existence of
sub-populations amongst the S0s: bright S0s, the ``real'' intermediate
class between E and Sa galaxies; and faint S0s, many of which could be
miss-classified as faint Es if viewed close to face-on.  The works of
Nieto et al. (1991) and J\/orgensen \& Franx (1994) support this
hypothesis, pointing out the similarity between disky [and thus faint
(Kissler-Patig 1997)] E and S0 galaxies, based on their isophotal
central shapes. Also, Graham et al. (1998), in a study of the
extended stellar kinematics of elliptical galaxies in the Fornax
Cluster, found that five of the galaxies are in fact rotationally
supported systems, suggesting that they could be misclassified S0
galaxies. 

Studies of the stellar populations in cluster galaxies (Kuntschner
2000, Smail et al. 2001) also support the idea of a dichotomy between
low- and high-luminosity S0s: bright S0s are old and coeval with E
galaxies, while faint members present younger central ages, indicating
more recent star formation episodes.  Furthermore, Poggianti et
al. (2001) examined the star-formation history of early-type galaxies
in the Coma Cluster, and found that $\sim40\%$ of the S0 population
seemed to have experienced a star-formation event during the last few
billion years, a phenomenon which is absent in their sample of elliptical
galaxies. Thus, it has been proposed that faint S0s could be the
descendants of the post-starburst galaxies found in high redshift
clusters. The work of Mehlert et al. (2000, 2003) in early type
galaxies in the Coma Cluster confirms the dichotomy found by Poggianti
et al. between old and young lenticulars; however, the high
alpha-element ratios found in the latter seems to argue against the
occurrence of recent star formation: the authors suggest that the
strong Balmer line indices measured in apparently ``young'' S0s could
actually be produced by unusually blue horizontal branches rather than
by young stellar populations.  Clearly, more work remains to be done
in the study of stellar populations to interpret the physical
significance of the two apparently-distinct types of S0.

The formation of lenticular galaxies has also been the subject of
numerical simulations.  The work of Shioya et al. (2004) on ``red
H$\delta$-strong'' galaxies also suggests two different evolutionary
paths for S0s, each one able to match different spectral features in
different galaxies: a ``truncation'' scenario, in which the star
formation is stopped and followed by passive evolution, and a
``starburst'' one, in which a relatively recent and short star
formation event precedes the cessation of star formation.  On the
other hand, Christlein \& Zabludoff (2004) found that simulations
based on a fading stellar population could not match the observed
luminosity distribution of galaxies with the larger
bulge-to-total-light ratios ($B/T$) typical of S0s.  They therefore
advocated a ``bulge enhancement'' model, where disk fading is
accompanied by an increase in the luminosity of the bulge; this model
seems to match the observations over a wide range of $B/T$.  Although
this work still faces the problem of the difficulty of reliably
identifying S0 galaxies, it does raise the interesting possibility
that, at least in some cases, the cessation of star formation in the
disk of the parent spiral galaxy could be preceded by a
centrally-concentrated episode of star formation, thus enhancing the
bulge luminosity. In this context, it is interesting that Moss \&
Whittle (2000) found spiral galaxies in clusters have more central
star formation than their field counterparts.

The extended stellar kinematics of S0 galaxies have been studied
mostly in conjunction with those of ellipticals, and often as one
single class of objects, generically termed ``early-type galaxies.''
For instance, D'Onofrio et al. (1995) studied a sample of 15
early-type galaxies in the Fornax Cluster, and did not find major
differences between Es and S0s other than stronger rotational support
and higher projected ellipticities for the latter. In a sample of 35 E
and S0 galaxies in the Coma cluster, Mehlert et al. (2000, 2003) found
that elliptical galaxies have, on average, slightly higher velocity
dispersions than S0s, as is also apparent from the velocity dispersion
profiles presented in D'Onofrio et al. (1995).  Although these
differences could be real, they may also at least in part arise from
the selection effects that render S0s more reliably identified when
close to edge-on.

With current techniques using integral-field units and high quality
spectra, it is possible to examine stellar-kinematic substructure in
search of further clues as to how these systems form.  The work by
Emsellem et al. (2004), for example, revealed that
kinematically-decoupled components, bars and misalignments between
photometric and kinematic axes seems to be present in both Es and
S0s.  There are also instances of even more extreme kinematic
substructure such as counter-rotating co-spatial stellar disks in S0s
(Rubin, Graham \& Kenney 1992), but these seem to be very rare
(Merrifield, Fisher \& Kuijken 1996).  This rarity is something of a
surprise, as counter-rotating gas is relatively common in S0s
(Bertola, Buson \& Zeilinger 1992), which led to the suggestion that
it might be quite common for S0s to be enhanced by the kind of minor
mergers likely responsible for this phenomenon.  These observations
can only be reconciled if there is some mechanism for inhibiting star
formation in such counter-rotating material, but the situation is
clearly quite complex.  

One last approach to understanding the origins of S0 galaxies lies in
their scaling relations.  In particular, there are a handful of
studies of the Tully--Fisher relation for S0 galaxies. Using a sample
of field S0 galaxies, Neistein et al. (1999) found a much larger
scatter in this relation between luminosity and rotation speed than is
found in spirals.  Hinz et al. (2001, 2003) found a similar result for
cluster S0s.  The large scatter would tend to indicate a rather stochastic
history for these systems rather than any simple deterministic
evolutionary path from spirals.  By contrast, Mathieu, Merrifield \&
Kuijken (2002) selected a sample of S0s with relatively small bulges
and found a Tully--Fisher relation with a small scatter but an offset
to fainter luminosities relative to the spiral relation, suggesting
a relatively simple fading of spirals into S0s for this sub-sample of
systems.  

The above discussion illustrates the wide range of techniques that
have been applied to trying to understand the nature of S0s, and the
rather contradictory results that have emerged.  There are certainly
indications of a dichotomy between faint and bright S0s, but it is
still unclear which observables best characterize this distinction,
and how those observables might translate into differences in the
evolutionary history of the two types.  Much of the difficulty arises
from the heterogeneous nature of the data that have been used in these
studies.  In some cases, the data come from objects in a range of
ill-defined differing environments, while in others the issues are
more to do with the varying quality of the observations.  

To address these issues, we have obtained very high signal-to-noise
ratio long-slit spectroscopy of 9 S0 galaxies in the Fornax Cluster.
The data are uniform in quality and probe in detail a single
environment, removing these variables from the problem.  However, the
sample does include both faint and bright examples of S0s, allowing us
to investigate whether there are systematic differences between these
systems in anything other than luminosity, and hence address the
question of whether they formed in different ways.  In this paper, we
present the details of the sample, the observations, the data
reduction, the resulting kinematics, and a qualitative look at the
results.  In subsequent papers, we will delve deeper into the
quantitative information that these data provide, looking at their
Tully--Fisher relation and the nature of their stellar populations out
to large radii.

The layout of this paper is as follows: Section~\ref{sec:sample}
describes the sample and observations; Section~\ref{sec:red} explains
the basic data reduction; Section~\ref{sec:kin} describes the
extraction of the kinematics; Section~\ref{sec:results} presents the
resulting kinematics and comparison to previous data;
Section~\ref{sec:vcirc} extracts the rotation curves from these data;
and Section~\ref{sec:conc} contains a summary.

\section[]{The sample and observations}\label{sec:sample}
The sample was selected from galaxies in the Fornax Cluster classified
as S0s by the NASA/IPAC Extragalactic Database.  They were selected to
span a wide range of luminosities ($-22.3<M_{B}<-17.3$), and to be
sufficiently inclined to measure rotations along their major axes.
The basic properties of the resulting sample of S0s are presented in
Table~1.  

The necessary observations of the major axes of these galaxies were
made in service mode between 2002 October 2 and 2003 February 24 at
the 8.2m Antu/VLT using the FORS2 instrument in long-slit spectroscopy
mode; exposure times and dates are provided in Table~1. 
Spectrophotometric standard stars were observed each night, and we
also obtained spectra of stars with a range of spectral classes to act
as velocity templates in the kinematic analysis; these objects are
listed in Table~2.  During the observations, the seeing varied from
0.75 to 1.48 arcsec FWHM, which is more than adequate for the study of
these large objects.

The detector in FORS2 comprises two 2k$\times$4k MIT CCDs, with a
pixel size of $15\times15\mu{\rm m}^2$. The standard resolution collimator
and the unbinned readout mode were used, yielding a scale of
$0.125^{\prime\prime}$/pixel. The spectrograph slit was set to
$0.5^{\prime\prime}$ wide and covered $6.8^\prime$ in length. The
GRIS$1400V+18$ grism was used, providing a dispersion of
$0.318$\AA/pixel and covering the 4560\AA~$ \le \lambda \le
5860$\AA\ wavelength range.  This set-up provided a spectral
resolution, as measured from the FWHM of the arc lines, of $\approx 4$ pixels
(or 1.12 \AA), which translates into a velocity resolution of $73.3\,$km$\,$s$^{-1}$
FWHM (or $31.0\,$km$\,$s$^{-1}$ in terms of the velocity dispersion). The CCD
was read out at $200\,$kHz, which is twice the normal speed used for
spectroscopy. The high readout speed was the only one available for unbinned
readout of the CCD.

\begin{table*}
 \begin{center}
  \caption{Sample of S0 galaxies in Fornax.}
  \begin{tabular}{@{}lcccccc@{}}
  \hline
   Name     & RA  & DEC & $B_{\rm T}^{*}$ & Diameter$^{*}$ & Exp.\ Time & Date\\
            &     &     &       &   [$^\prime$]    &  [sec]    \\
 \hline
 NGC\,1316 & 03 22 41 & $-$37 12 30 & 9.40 & 11.0 & $3\times1200$ & 13 Oct 2002\\
 NGC\,1380 & 03 36 27 & $-$34 58 34 & 10.9 &  4.8 & $2\times1200$ & 24 Feb 2003\\
 NGC\,1381 & 03 36 31 & $-$35 17 43 & 12.4 &  2.7 & $2\times1600$ & 24 Feb 2003\\
 IC\,1963  & 03 35 30 & $-$34 26 51 & 12.9 &  2.6 & $2\times1600$ & 31 Jan 2003\\
 NGC\,1375 & 03 35 16 & $-$35 15 56 & 13.2 &  2.2 & $2\times1800$ & 28 Dec 2002\\
 NGC\,1380A& 03 36 47 & $-$34 44 23 & 13.3 &  2.4 & $2\times1700$ & 28 Dec 2002\\
 ESO\,358$-$G006 & 03 27 18 & $-$34 31 35 & 13.9 & 1.2 & $2\times2400$ & 14 Oct
 2002\\
 ESO\,358$-$G059 & 03 45 03 & $-$35 58 22 & 14.0 & 1.0 & $1\times2550$ & 8 Feb
 2003\\
 ESO\,359$-$G002 & 03 50 36 & $-$35 54 34 & 14.2 & 1.3 & $1\times2250$ & 26 Nov
 2002\\
  \hline
\end{tabular}\\
\end{center}
\footnotesize{$^{*}$ From RC3, de Vaucouleurs et al.\ (1991).}
\end{table*}

\section[]{Data Reduction}\label{sec:red}
 
The standard data reduction process was carried out using IRAF (Tody
1986, 1993). Bias subtraction, flat-fielding and cosmic ray removal
were applied to all science and calibration images. Bad pixel masks
were created by dividing two flatfield images of different exposure
times, and the bad pixel values were interpolated.  When more than one
exposure was obtained for each galaxy (see Table~1), the individual
spectral images were combined to maximise the signal-to-noise ratio
(S/N).  During the combination process the position of the galaxy
spectrum and the locations and widths of several sky lines were
checked; the match between the different exposures for each galaxy was
found to be excellent, so no further alignment was necessary.

After tracing the arc lines of the He--Ne lamp, a 2D correction was
applied to all the spectra in order to remove the geometric
distortions due to the instrument optics and CCD flatness issues. The
wavelength calibration was applied in the same step.  The residuals of
the wavelength fits were typically 0.1--0.2\AA. 

Next, we subtracted the night sky spectrum using the IRAF task {\tt
background}.  Potentially, residuals and uncertainties from sky
subtraction could bias the kinematic and line strength measurements
that we will make from these data, so we have looked quite closely at
how this step is implemented.  Given the large spatial coverage of the
data, there are sizeable regions that are free of galaxy light from
which we can define a sky spectrum, and we found that the resulting
sky-subtracted galaxy spectrum was almost completely insensitive to
how the sky region was chosen.  We have therefore opted for the
simplest choice of selecting sky regions, typically 100 pixels wide,
on each side of the galaxy and as far from it as possible within the
confines of the slit.  The only exception is the case of NGC\,1316,
the largest and brightest galaxy, where only one sky region was used
because the galaxy centre was placed close to one end of the slit in
order to maximise the spatial coverage.  Contamination by scattered
light was not found to be a significant issue: in all cases, there
were no nearby bright sources, and the relatively low surface
brightnesses of the systems reduces the likelihood of misplaced light
from the galaxies themselves.

A periodic, square wave noise pattern or `hum' was found to be present
along the dispersion direction in all the spectra. The amplitude of
this hum varied from 3 to 6 counts, and its frequency, 0.077
cycles/pixel ($15\,$kHz), was reasonably constant, with small
deviations due mainly to previous data processing. Because this
pattern could partially mask the spectral features at low S/N, we
subtracted it using a Fourier technique (Brown 2001).  Ideally, one
would apply this technique to the raw data frames, but its low level
meant that it could only be reliable identified and removed after
processing to subtract the sky.  In practice, inspection of the
resulting cleaned images showed that this processing had no adverse
impact on our ability to remove this source of systematic noise.

To correct by the effect of atmospheric extinction, we used the IRAF
task {\tt setairmass} to define the effective airmass for each
spectrum. Then, we applied the extinction correction (task {\tt
calibrate}) using the appropriate atmospheric extinction table. The
spectra were subsequently flux calibrated using the set of
spectrophotometric standard stars listed in Table~2, which allowed us
to transform the observed counts into relative spectral fluxes as a
function of wavelength.  The sensitivity functions as derived from the
different stars varied by typically $\sim 1$\%, reaching $\sim 3$\%
towards the edges of the wavelength range, so we created a single
function from all of them and can be reasonably confident that the
resulting internal relative flux calibration is good to $\sim 1\%$.

This analysis was applied both to the galaxy spectra and to the
spectra of the 10 template stars (listed in Table~2), which will be
used to model the spectra of the galaxies during the extraction of the
kinematics.

\begin{table}
 \centering
 \caption{Spectrophotometric (S) and Template (T) Stars.}
 \begin{tabular}{@{}lcc@{}}
  \hline
   Name  & T/S   & Spectral Class \\
  \hline
  BD-01-0306 & T & G1 V \\
  HD1461     & T & G0 IV \\
  HD7565    & T & K2 III-IV \\
  HD16784     & T & G0 V \\
  HD18234     & T & K0 \\
  HD19170     & T & K2 \\
  HD21197    & T & K5 V \\
  HD23261    & T & G5 \\
  HD36395    & T & M1.5 V \\
  HD61606    & T & K2 V \\
  \hline
  LTT1020    & S & --   \\
  LTT1788    & S & --   \\
  LTT3218    & S & --   \\
  HZ4        & S & --   \\
  GD71       & S & --   \\
  Feige67    & S & --   \\
  \hline
\end{tabular}
\end{table}

The calibrated two-dimensional galaxy spectra were binned along the
spatial direction so as to generate one-dimensional spectra from
different radii with comparable S/Ns.  We started at the centre of the
each galaxy using bins with a S/N per \AA ngstrom of 53 (or 30 per
pixel). When we reached the outer parts, where no more bins with that
S/N ratio could be built, we reduced the S/N of each bin to 30 per \AA
ngstrom, then to 15, and finally to 10. The data were also re-binned
on to a logarithmic wavelength scale in the dispersion direction to
enable the kinematic measurements.

\section[]{Extraction of the kinematics}\label{sec:kin}

In the present study, the kinematic properties of the galaxies were
derived using the Penalised Pixel Fitting method (pPXF, Cappellari \& Emsellem 2004),
which is a parametric technique based on maximum penalised likelihood.
Since the resulting kinematic parameters are so central to this work,
we first describe the workings of this method as they apply to the
current data set, and the tests that we undertook to check the
reliability of the resulting parameters.

The parametric method used by pPXF models the line-of-sight velocity
distribution (LOSVD) as a Gaussian plus a series of Gauss-Hermite
polynomials. The fits return the mean line-of-sight velocity, $V_{\rm
LOS}$, and the velocity dispersion, $\sigma$, from the Gaussian,
together with the $h_3$ and $h_4$ coefficients from the
polynomials. These two coefficients are related, respectively, to the
skewness and the kurtosis of the LOSVD, which are higher moments
associated with the asymmetric and symmetric departures of the LOSVD
from a Gaussian.

\begin{figure}
\begin{center}
\includegraphics[scale=0.5]{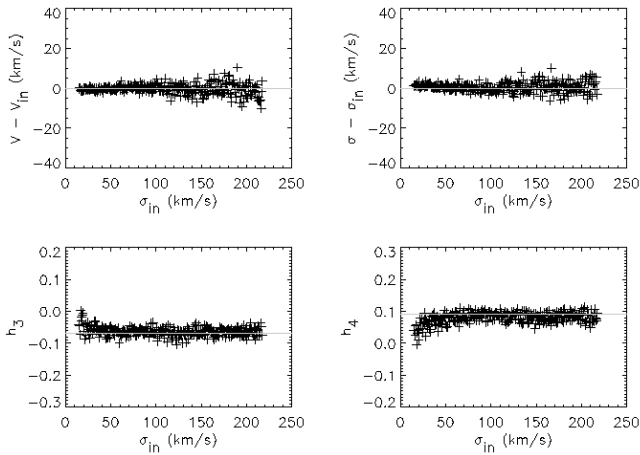}
\end{center}
\caption{ Sample 1000 pPXF simulations carried out in order to find
the optimum value of the $\lambda$ parameter for this dataset. In this
particular case, the simulated spectra have S/N $= 53$ and $\lambda$
has been set to 0.3. The model LOSVD has $V_{\rm in}=0$, $h_{3\rm in}
= -0.07$, $h_{4\rm in} = 0.09$ and the derived values of $V_{\rm out}$,
$h_{3\rm out}$ and $h_{4\rm out}$ are shown for different values of
$\sigma_{\rm in}$.}\label{fig:snsim}
\end{figure}

This software works in pixel space, finding the combination of template spectra which,
convolved with an appropriate LOSVD, best reproduces the galaxy spectrum in
each bin. A subroutine of the program fits the continuum, using
Legendre polynomials of order 4 (in this case), and divides the spectra
by the fit to remove any possible low-frequency mismatches between
the galaxy spectra and the model. The method carries out a
pixel-by-pixel minimisation of the residuals (quantified by $\chi^2$),
adding an adjustable penalty term to the $\chi^2$ to bias the
resulting LOSVD towards a Gaussian shape when the higher moments $h_3$
and $h_4$ are unconstrained by the data. Thus, a deviation $D$ of the
LOSVD from a Gaussian shape will be accepted as an improvement of the
fit only if it is able to decrease the residuals by an amount
\begin{center}
\begin{equation}\label{eq:eps}
\chi^2_p = \chi^2(1+\lambda^2\cdot D^2).
\end{equation}
\end{center}
The $\lambda$ parameter defines the threshold between a deviation
considered as an improvement of the pure Gaussian fit, and one
considered as due to noise, and thus rejected.  As Cappellari \& Emsellem (2004) suggest,
the value of $\lambda$ has to be determined for each particular
dataset because the ability of the software to find the correct LOSVD
parameters (especially the higher moments) will depend critically on
the S/N of the data and other properties of the particular spectra
such as the spectral range covered.  To estimate the optimum parameter
for our study, we performed Monte-Carlo simulations in which we
convolved the template spectra with a LOSVD whose shape was specified
by model parameters $V_{\rm in}$, $h_{3 \rm in}$ and $h_{4 \rm in}$,
and whose width was allowed to vary such that $0 < \sigma_{\rm in} <
250\,$km$\,$s$^{-1}$.  These spectra, degraded to the S/N of the
various one-dimensional galaxy spectra, were then passed through the
pPXF software to recover estimates for the parameters, $V_{\rm out}$,
$\sigma_{\rm out}$, $h_{3 \rm out}$ and $h_{4 \rm out}$.  An example
of one such run is presented in Figure~\ref{fig:snsim}.  After
extensive testing using a variety of input parameters and S/N,
we found that $\lambda=0.3$ provided an optimum value for the spectra
in the current dataset, yielding values that minimized both errors and
bias in the results.

As is apparent from Figure~\ref{fig:snsim}, even at this fairly modest
value for the $\lambda$ parameter, a degree of bias appears in the
answers at low dispersion, particularly in the derived values of
$\sigma$ and $h_4$.  This kind of bias, particularly prevalent in such
even moments, is a well-known phenomenon (e.g. Cappellari \& Emsellem
2004).  Clearly, we must understand and possibly correct for this
effect before interpreting the calculated parameters, so we have
performed a range of tests using our adopted value of $\lambda$ but
allowing other parameters to vary.  The conclusion of this study is
that the bias in $\sigma$ is small but well-defined at a maximum level
of $\pm2$-$3\,$km$\,$s$^{-1}$; we corrected the results for this bias,
but it makes no substantive difference.  The case for $h_4$ is a
little more disturbing, as this parameter, particularly in galaxies
with low values of $\sigma$, can suffer significant bias, amounting to
offsets as high as 0.4 in the most extreme cases.  The uncertain
nature of this bias means that we have not corrected for it explicitly
in our final derived value, but we would caution against interpreting
the $h_4$ values as too precise, particularly from spectra where S/N
is low and where $\sigma$ is small.  Nonetheless, this study does
demonstrate that the principal parameters of interest, $V_{\rm LOS}$
and $\sigma$, are robustly extracted from spectra of the quality that
we present here, irrespective of the detailed shape of the LOSVD as
specified by $h_3$ and $h_4$.

To estimate the errors on the derived parameters for each
one-dimensional galaxy spectrum, we ran a further series of Monte
Carlo simulations.  We modelled each such spectrum using the corresponding
best combination of template
stars, broadened by the best-fit LOSVD, and degraded to the S/N of the
galaxy data.  These model spectra were then passed through the pPXF
software to measure the distribution of the derived parameters, and
hence a measure of the uncertainty in a single spectrum.  The
dispersion in the parameters derived in each Monte Carlo simulation is
quoted as the error bar on the corresponding parameter derived for the
real galaxy spectrum.

Before we can present the final data set of kinematic parameters
versus distance along the major axes for these S0 galaxies, the only
remaining issue is to define the zero-points of velocity and position
for each galaxy.  The velocity zero-point was simply set to the median
value of $V_{\rm LOS}$ determined along the slit for each galaxy.  For
the spatial zero-point -- the galaxy's centre -- the simplest
definition would just be the point along the slit that admitted the
greatest amount of light.  However, this definition could well pick up
on an off-centre localized feature that would then distort the
over-all kinematic structure of each galaxy that is the primary goal
of this study.  Instead, we adopted as the spatial origin the position
along the slit that renders the plot of $V_{\rm LOS}$ versus radius as
close to anti-symmetric as possible.  The only case where this method
was not applied was NGC~1316; the large size of this system meant that
we placed the slit primarily to cover only one side of the galaxy, so
we could not subtract the median velocity along the slit nor match up
the two sides of the rotation curve to find the kinematic centre.  For
this galaxy, we used the point along the slit at which we derived the
largest value of $\sigma$ to define the spatial location of the
galaxy's centre, and then used the value of $V_{\rm LOS}$ at this
point to specify the galaxy's systemic velocity.

\section{Results and comparisons}\label{sec:results}
The values for the kinematic parameters $V_{\rm LOS}$, $\sigma$, $h_3$
and $h_4$ versus radius, as derived using the analysis set out in
Section~\ref{sec:kin}, are presented in Appendix~\ref{app:data}.

A number of these galaxies have been studied previously with lower quality
spectra or less sophisticated spectral analyses, so as well as
presenting our own results we will also compare the data to these
published results.  The most extensive study for comparison is that of
D'Onofrio et al.\ (1995), who obtained major-axis kinematics ($V_{\rm
LOS}$ and $\sigma$) for six galaxies in common with the current
sample.  One complication in comparing their data with the new
observations is that they presented the kinematics folded about the
centre of each galaxy, leading to an ambiguity in the actual spatial
location of each measurement.  We therefore cannot match up individual
features that occur on one side of a galaxy, but can use these data to
examine the overall profile.  The other limitation of their data was
its relatively low spectral resolution ($\sim 60\,{\rm km}\,{\rm
s}^{-1}$ in velocity dispersion), which, as we will see, introduces a significant bias into
the kinematics.  Further long-slit data for single galaxies in the
current sample can be found in Graham et al. (1998) and Longhetti et
al.\ (1998), so we also include these data in our comparison.  There
are also a number of measurements of the central velocity dispersion
that provide a single comparison point.  Kuntschner (2000) obtained
central dispersions for all the galaxies in this sample, although he
warns that the relatively low spectral resolution of his data,
corresponding to a velocity dispersion of $\sim50\,{\rm km}\,{\rm
s}^{-1}$, mean that the measurements should only provide a rough guide
for the fainter galaxies.  Further measurements can be found in the
compilation of Bernardi et al. (2002), although the variety of sources
from which these data were drawn mean that their reliability is less
assured than in the other measurements.

\subsection{Results for individual galaxies}
In the following paragraphs we give a brief description of the observed
kinematics for each individual galaxy in the sample. We discuss these
result and compare them with the previous studies described above.  

\begin{figure*}
\begin{center}
\includegraphics[scale=0.4]{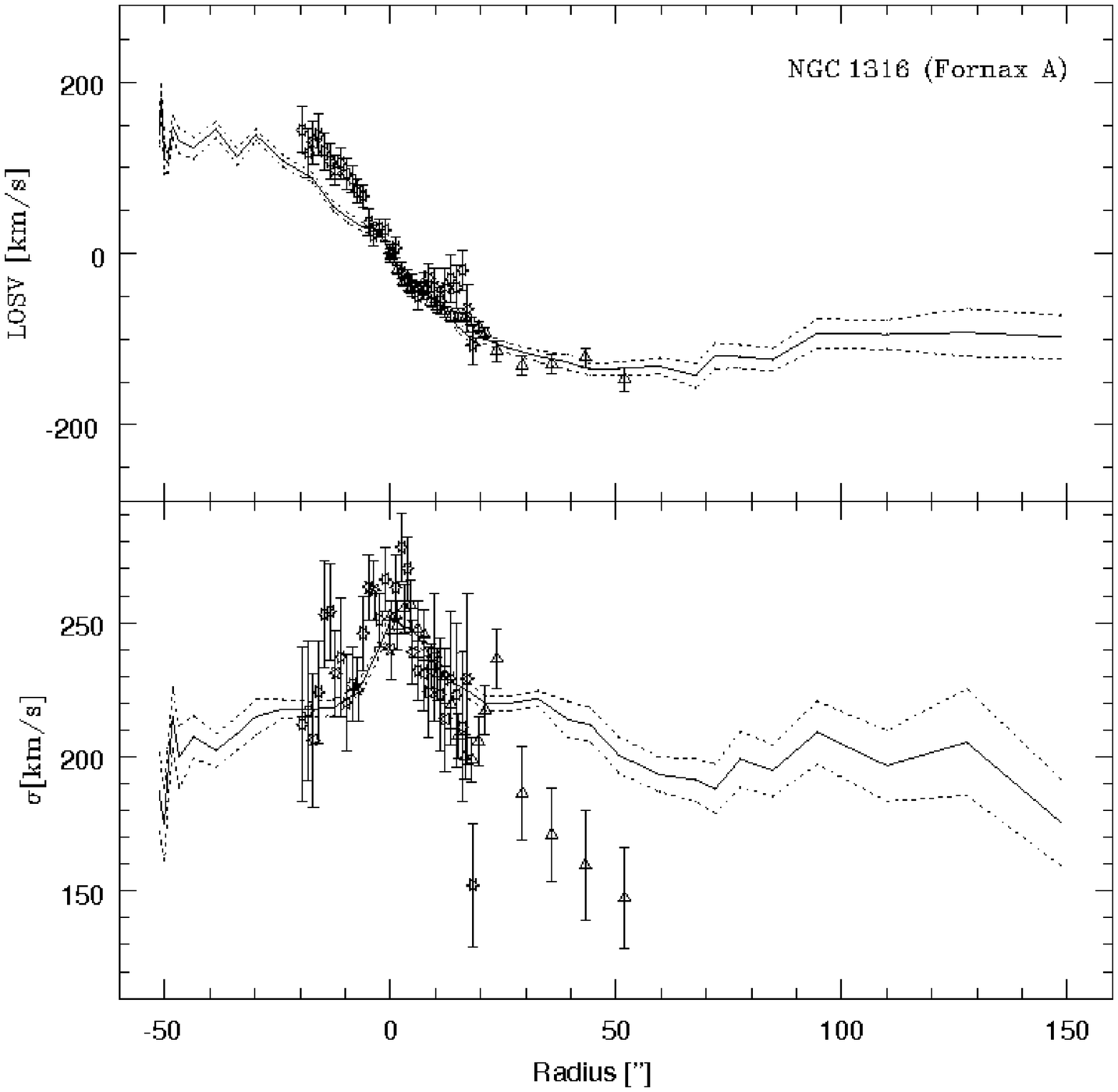}
\includegraphics[scale=0.4]{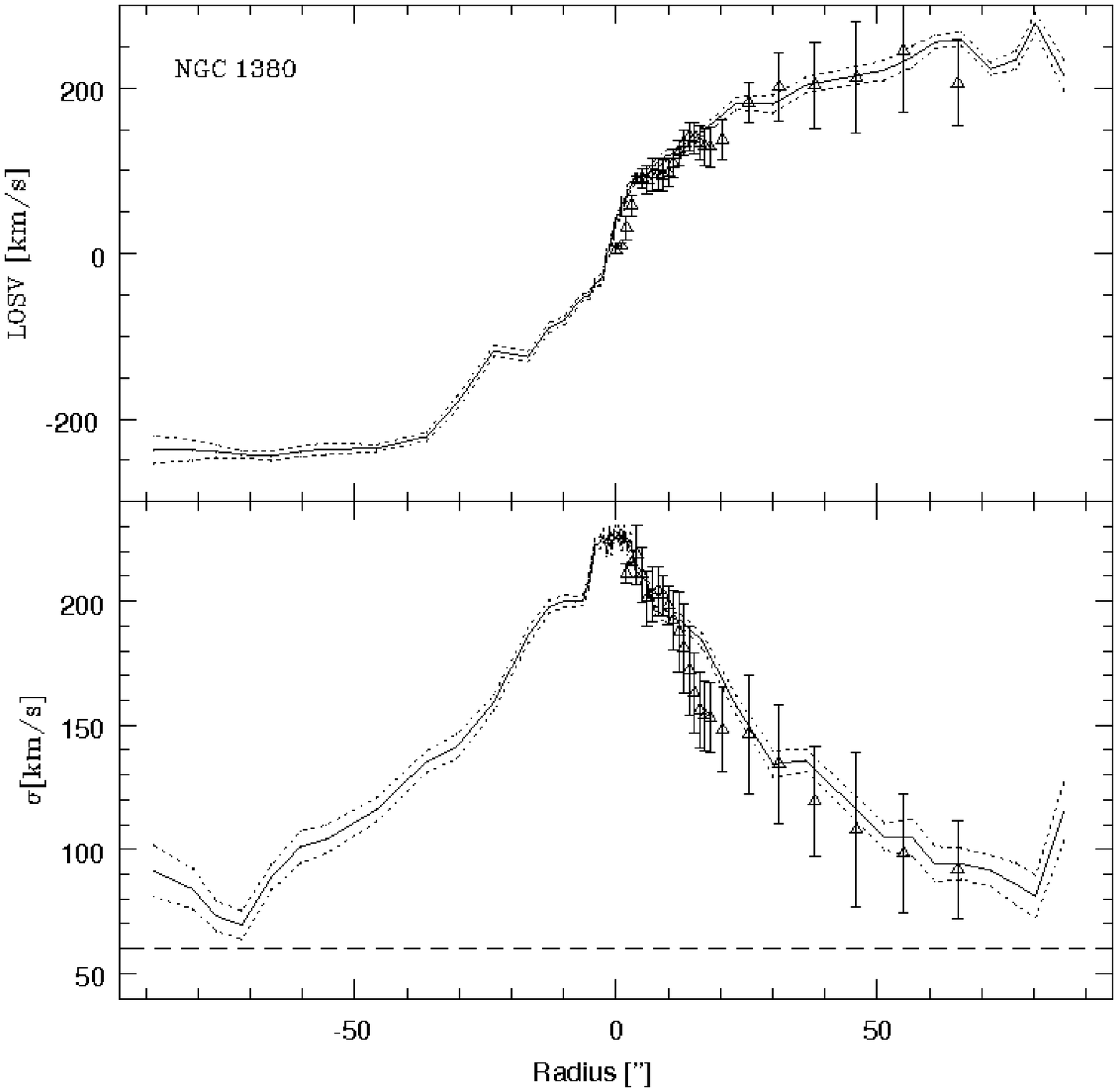}
\includegraphics[scale=0.4]{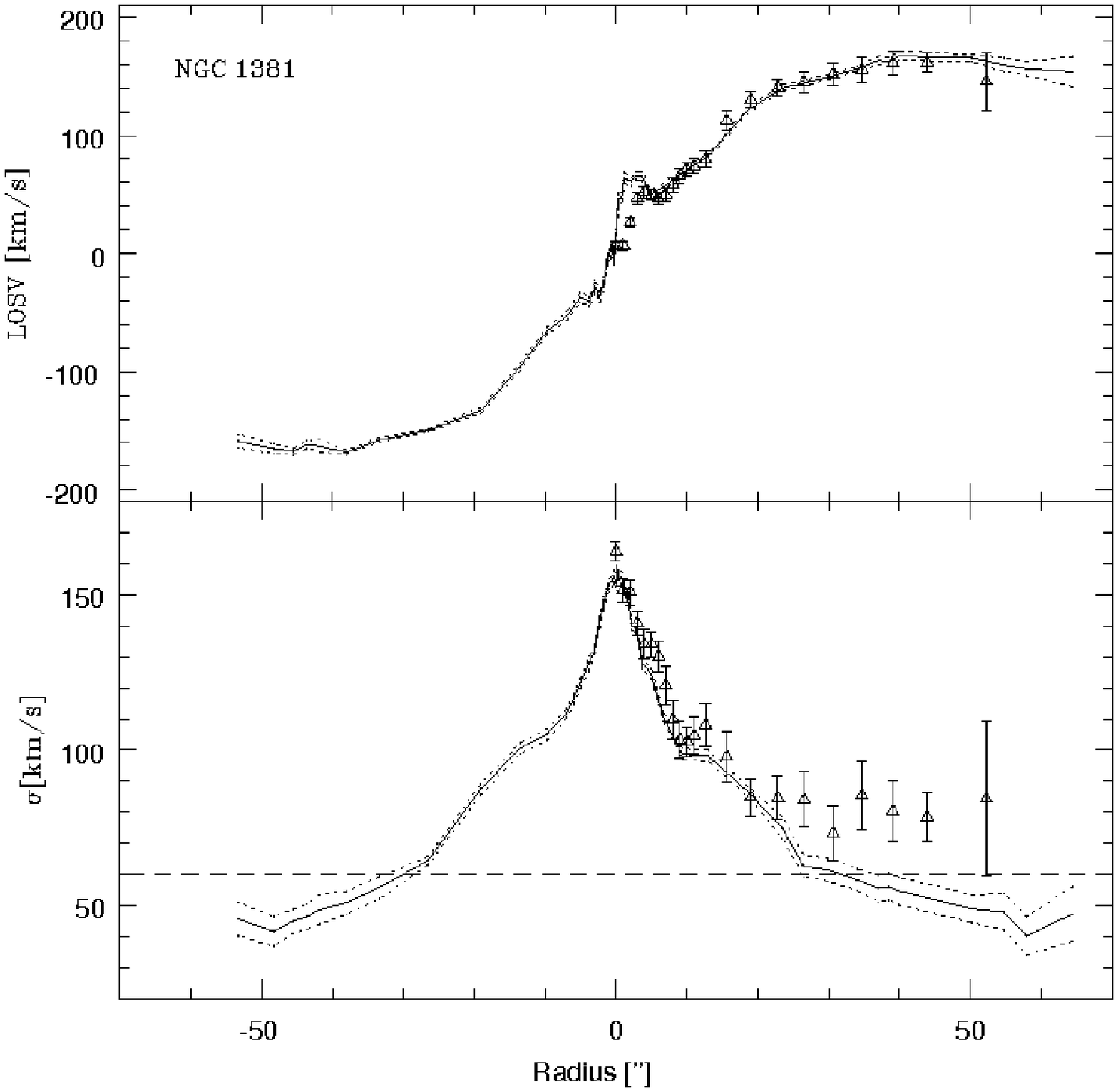}
\includegraphics[scale=0.4]{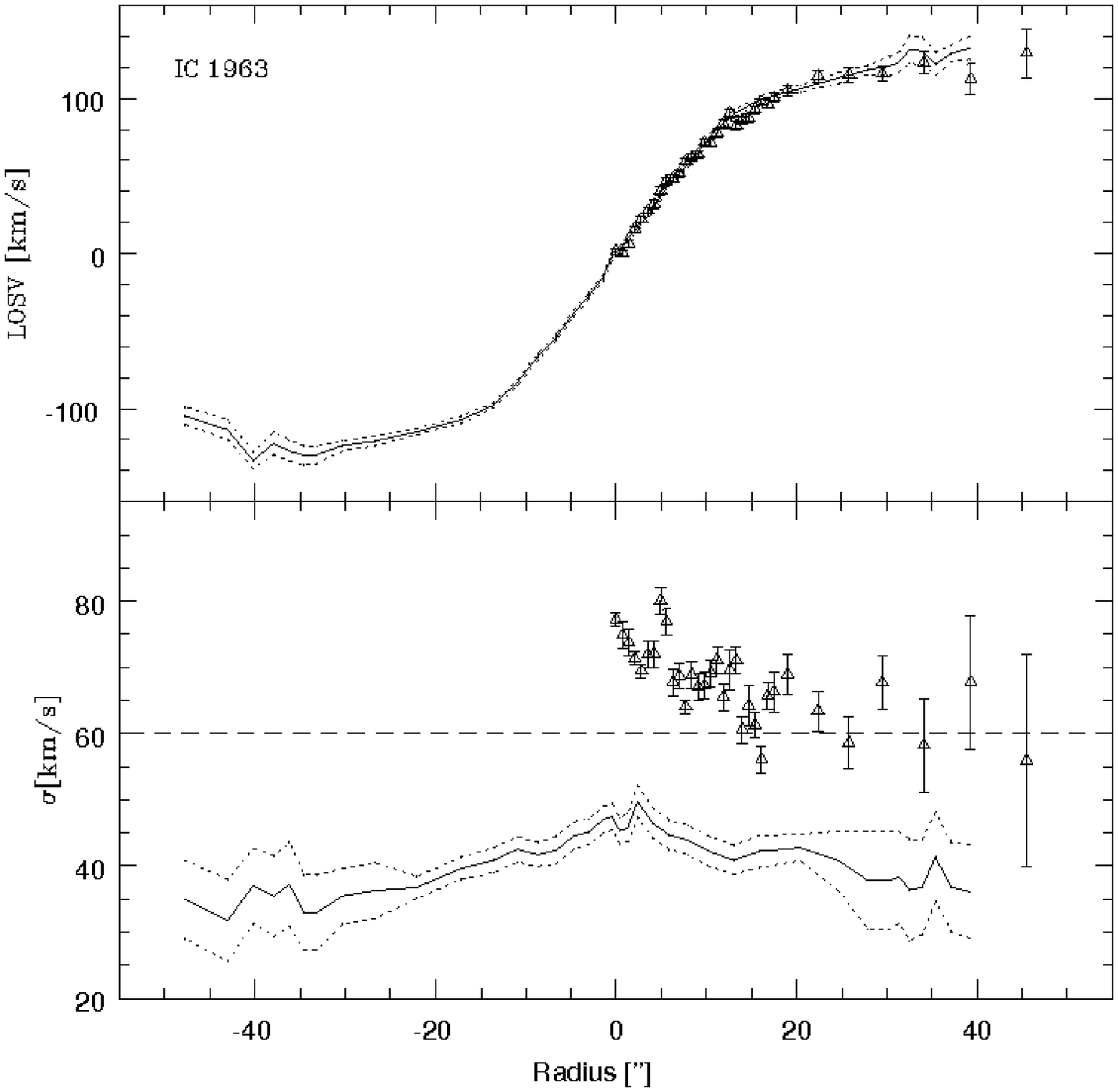}
\end{center}
\caption{\small Comparison between the new results for mean velocity,
  $V_{\rm LOS}$, and velocity dispersion, $\sigma$, and kinematics
  from the literature for NGC\,1316, NGC\,1380, NGC\,1381 and
  IC\,1963.  The new results are shown as a continuous line with
  $1\sigma$ errors as short dashed lines.  The D'Onofrio et al.\ (1995)
  data are shown as triangles, the Longhetti et al.\ (1998) data are
  shown as stars, and the Graham et al.\ (1998) data are shown as
  squares.  Horizontal dashed lines in the velocity dispersion plots
  indicate the resolution limit of the archival data.}\label{fig:comp1}
\end{figure*}

\begin{figure*}
\begin{center}
\includegraphics[scale=0.4]{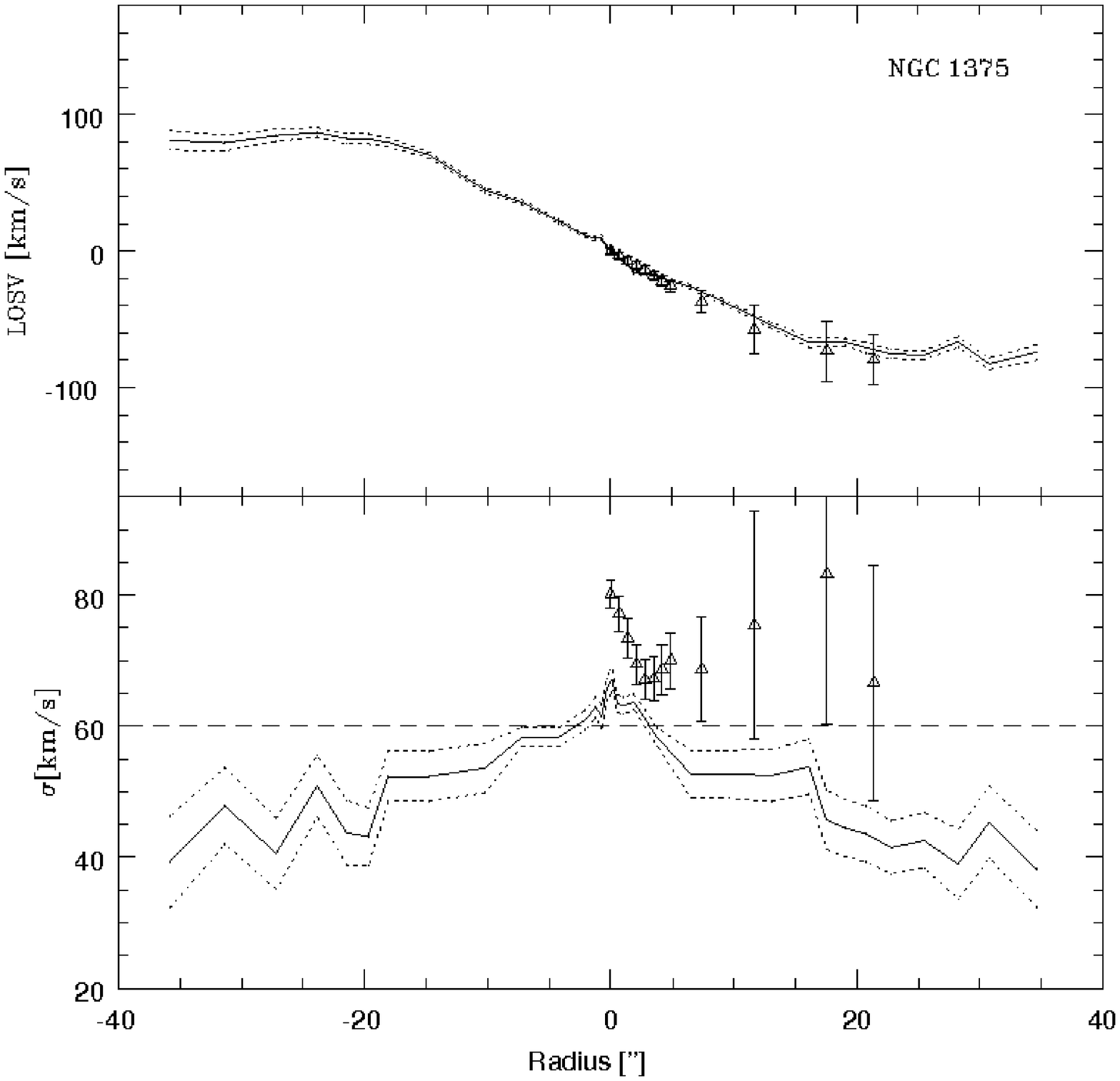}
\includegraphics[scale=0.4]{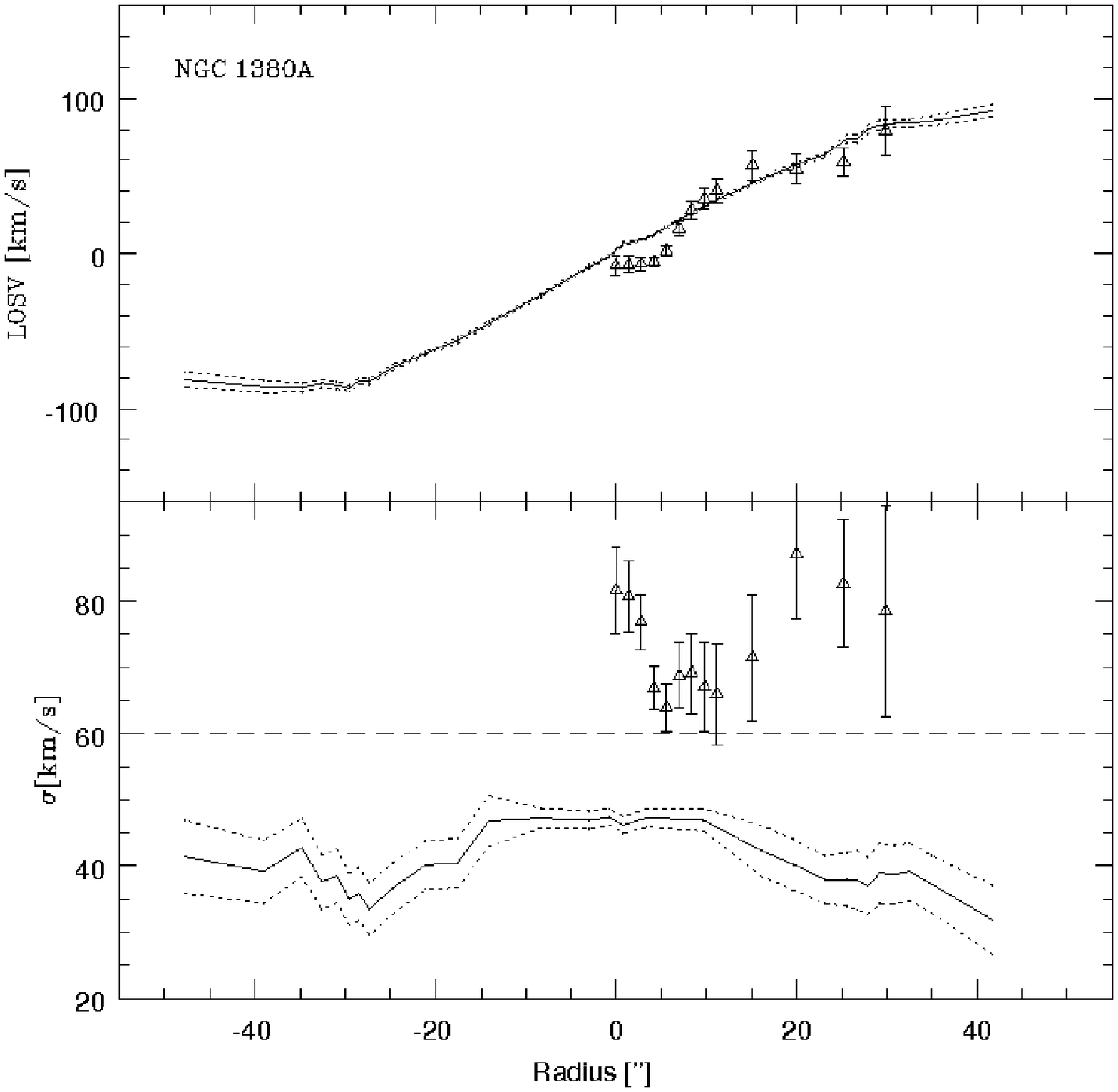}
\includegraphics[scale=0.4]{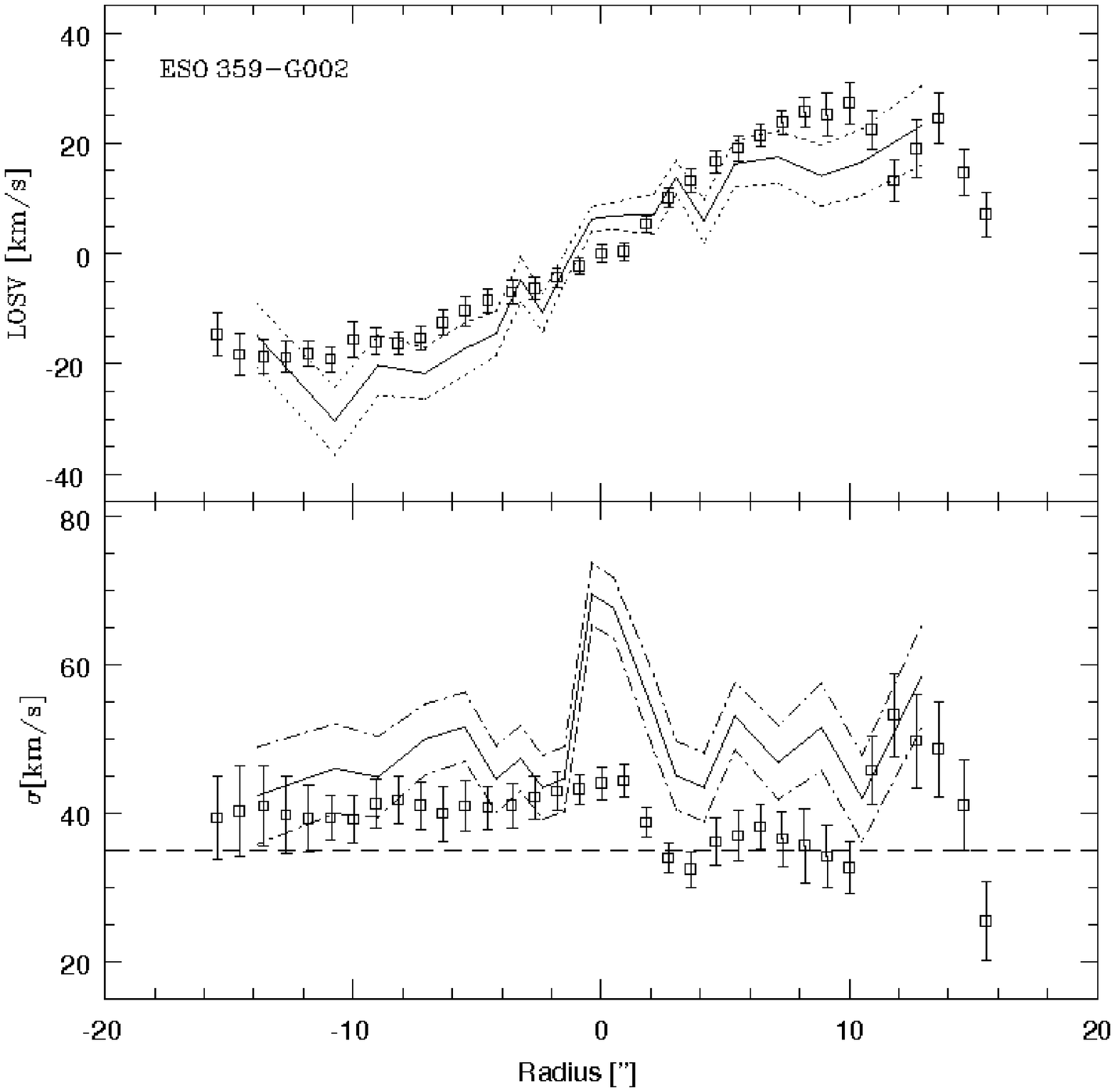}
\end{center}
\caption{As for Figure~\ref{fig:comp1}, for NGC\,1375,
NGC\,1380A and ESO\,359-G002}\label{fig:comp2}
\end{figure*}

\subsubsection{NGC\,1316 (FORNAX A)}
For this well-known merger remnant, Figure~\ref{fig:comp1} reveals
generally good agreement with the published results, although the new
data reach to very much larger radii.  The new data indicate a rather
higher dispersion than the points at the largest radii of the previous
measurements, but since these points were derived from the lowest S/N
data in the older observations, the conflict is probably not
significant.  The more extended velocity dispersion profile from the
new data for the first time reveals a seemingly-distinct feature of
enhanced dispersion within the central $15\,{\rm arcsec}$. In this
galaxy, as might be expected in a merger, random motions dominate out
to large radii in the new data, bringing into question whether this
galaxy should really be classified as an S0 system.  

\subsubsection{NGC\,1380}
Figure~\ref{fig:comp1} shows that the new data agree extremely well
with the previous observations, but with much improved error bounds.
The new data with their smaller errors and unfolded presentation give
some indication of localized substructure that differs between the two
sides of the galaxy.  An enhancement in velocity dispersion within the
central $\sim 7\,{\rm arcsec}$, first suggested by D'Onofrio et al.\
(1995), is strongly confirmed with the new data.  It is also notable
from Figure~\ref{fig:1380dat} that this region also shows structure in
the $h_3$ skewness measure, indicative of a distinct kinematic
component.  Interestingly, this feature coincides with photometric
evidence for a distinct component which comprises a strong variation
in isophotal ellipticity and position angle (Caon et al. 1994).

\subsubsection{NGC\,1381}
Figure~\ref{fig:comp1} reveals a distinctive feature in this galaxy's
rotational motion which was not apparent in the previous data.  There
is a correspondingly very strong signal in the $h_3$ profile shown in
Figure~\ref{fig:1381dat}.  Such kinematic features are characteristic
of a bar viewed edge-on (Bureau \& Athanassoula 2005), so we would
seem to have identified this edge-on galaxy, previously designated as
unbarred, as a barred system.  The new velocity dispersion profile
diverges at large radii from the D'Onofrio et al.\ (1995) data, but
this would seem to be because the latter has been artificially
enhanced in regions where the dispersion approaches the resolution
limit of the older data.

\subsubsection{IC\,1963}
Figure~\ref{fig:comp1} shows very good agreement between new
measurements of mean streaming velocity and those in D'Onofrio et al.\
(1995) for this rather nondescript S0 system.  The velocity dispersion
measurements do not agree at all well, but once again this conflict
can be attributed to the inadequate spectral resolution of the older
data.

\subsubsection{NGC\,1375}
Once again, Figure~\ref{fig:comp2} shows that the measured rotational
motion for this galaxy agrees well with previous data, while the
disagreement with earlier velocity dispersion data seems to arise from
the limited spectral resolution of the older data.  Support for this
interpretation of the inconsistency comes from Kuntschner's (2000)
value for the central dispersion of $56 \pm 10\,$km$\,$s$^{-1}$, in
agreement with the new data.  A further issue with the earlier
analysis is revealed by Figure~\ref{fig:1375dat}, as the structure in
$h_3$ and $h_4$ at small radii would indicate that there is more
information in these dynamics than would be revealed by the earlier
Gaussian fit.  Once again, this structure indicates the presence of a
distinct kinematic component such as a bar.  Photometric support for
this conclusion comes from the work by Lorenz et al. (1993), who found
that NGC\,1375 possesses a complex structure characterised by boxy
isophotes within a radius of $15\,{\rm arcsec}$; such boxy isophotes
are often associated with the kinematic signature of an edge-on bar
(Kuijken \& Merrifield 1995).

\subsubsection{NGC\,1380A}
The well-constrained measurements of rotational motion for this galaxy
are fairly featureless, and, as Figure~\ref{fig:comp2} shows, do not
reproduce the rather strange profile found by D'Onofrio et al.\
(1995).  The similar absence of peculiarities in the $h_3$ and $h_4$
profiles shown in Figure~\ref{fig:1380Adat} suggest that the velocity
structure in this galaxy is quite simple, so the smooth mean
velocity curve is not unexpected.  Once again, the new measurements of
velocity dispersion lie systematically well below those in D'Onofrio
et al.\ (1995), presumably due to the spectral resolution of the
latter; Kuntschner's (2000) published central dispersion of $55 \pm
9\,{\rm km}\,{\rm s}^{-1}$ fits rather better with the new
measurements, but may also be running into resolution issues.

\subsubsection{ESO\,359-G002}
As Figure~\ref{fig:comp2} shows, this fainter lenticular galaxy has a
slowly rising mean velocity curve, in good agreement with the previous
data from Graham et al. (1998) and despite being below our estimated spectral
resolution. The previous velocity dispersion measurements are in reasonable
agreement with the new data, but seem to lie systematically lower.  In
particular, the sharp central peak in velocity dispersion, which is
accompanies by a similar structure in $h_4$ (see Figure~\ref{fig:G002dat}),
was not seen in the earlier data. In this case both, the velocity curve and
the dispersion profile should be viewed with some caution.

\subsubsection{ESO\,358-G006}
There are no previous studies of the extended kinematics for this
faint galaxy. Figure~\ref{fig:G006dat} reveals generic featureless
rotational motions that rise to a plateau at large radii, and a rather
strange velocity dispersion profile with a slight central depression
and a hint of a rising profile at large radii.  The existing values
for the central velocity dispersion of $58 \pm 8\,{\rm km}\,{\rm
s}^{-1}$ from Kuntschner (2000) and $46 \pm 3\,{\rm km}\,{\rm s}^{-1}$
(Bernadi et al. 2002) are in accord with the new dispersion
measurements.

\subsubsection{ESO\,358-G059}
The extended kinematics of this faint galaxy have not previously been
studied.  Figure~\ref{fig:G059dat} shows the rather strange kinematics
for this system.  It has a relatively well-behaved mean velocity
profile that rises to an amplitude of $\sim 50\,{\rm km}\,{\rm
s}^{-1}$ at $\sim 5\,{\rm arcsec}$, then declines slightly.  The
dispersion profile is remarkably flat in the region where the rotation
increases, then starts to rise when the rotational velocities decline.
The published values for the central velocity dispersion, $54\ \pm 9 {\rm km}\,{\rm s}^{-1}$ (Kuntschner 2000) and $46 \pm 3\,{\rm
km}\,{\rm s}^{-1}$ (Bernadi et al. 2002), are in reasonable agreement
with the new data.  One possible explanation for the enhanced
dispersion and depressed rotation at larger radii would be the
presence of a counter-rotating stellar disk (Rubin et al. 1992), but
for such a faint galaxy we clearly do not have data of sufficient
quality to test this hypothesis rigorously.

\section{Circular velocity calculation}\label{sec:vcirc}
In order to go beyond the kinematic measurements of
Section~\ref{sec:results} to place these observations in their
astrophysical context, we have to translate them into the related
intrinsic dynamical properties of the galaxies.  In particular, since
we ultimately want to explore the Tully--Fisher relation for these
galaxies, we need to derive the circular velocity as a function of
radius, $V_{\rm C}(R)$.  Although this quantity is clearly connected to the
observed mean velocity profile, $V_{\rm LOS}(R)$, the transformation
from one to the other is not trivial, so we describe the procedure we
have adopted in some detail.  

The first ingredient we need is the inclination, $i$, of the galaxy to
the line of sight, in order to determine the fraction of rotational
motion that will have been projected into the observed line-of-sight
mean velocity.  In principle, the inclination can be estimated by
simply measuring the flattening of isophotes at large radii, but this
approach is not advisable as it uses only the lowest S/N photometry.
Instead, we have carried out a full two-dimensional fit to images of
the galaxies using the GIM2D software (Simard et al.\ 2002).  We have
applied this software to the publicly-available 2MASS $K$-band images
of the objects in our sample (Jarrett et al. 2003).  The uniform
quality of these data means that we can study all of the galaxies in a
consistent manner, and the use of infrared light means that we are
tracing the bulk of the population in these systems, avoiding any
problems that might arise from localized star formation or dust
obscuration.  In addition to obtaining a best estimate for the
galaxies' inclinations, this analysis also returns estimates for the
parameters of a disk-plus-bulge model, including a bulge effective
radius, $R_{\rm e}$, and a disk scalelength, $R_{\rm exp}$.  As we
will see below, these photometric parameters are also useful in
determining the properties of the rotation curve, so their values are
presented in Table~\ref{tab:photpar}.

\begin{table*}
 \begin{center}
 \caption{Structural and other important parameters of S0 galaxies in
 the Fornax Cluster.}\label{tab:photpar}
  \begin{tabular}{@{}lccccccccc@{}}
  \hline
   Name & $i$ & $B/T$ & $R_{\rm e}$ & $R_{\rm exp}$ & S\`ersic & $R_{\rm LIM}$& $R_{\rm out}$ & $V_{\rm max}$ & $M_B$ \\
        & [${}^o$] &  &  ['']  & ['']    &   $n$      &     ['']  &    ['']
	\{$[R_e]$\}   &    [km$\,$s$^{-1}$] &        \\
    (1) & (2) &  (3)  &  (4)  &    (5)    &     (6)    &     (7)  &    (8)    &    (9)    &(10)  \\
 \hline
 NGC\,1316 &  $43.2^{\phantom{1}43.8}_{\phantom{1}42.4}$ & $0.58^{\phantom{1}0.62}_{\phantom{1}0.47}$ & $36.0^{\phantom{1}41.7}_{\phantom{1}24.8}$ & $49.4^{\phantom{1}52.7}_{\phantom{1}47.2}$ & $2.9^{\phantom{1}3.2}_{\phantom{1}2.3}$ & --  & 148.5 \{4.1\} & -- & -22.3 (0.1)\\
 \\
 NGC\,1380 &  $66.9^{\phantom{1}67.5}_{\phantom{1}66.5}$ & $0.58^{\phantom{1}0.59}_{\phantom{1}0.57}$ & $17.6^{\phantom{1}18.0}_{\phantom{1}17.3}$ & $36.3^{\phantom{1}37.3}_{\phantom{1}35.9}$ & $3.3^{\phantom{1}3.4}_{\phantom{1}3.3}$ & 35.7 & 88.6 \{5.0\} & 309.6 (26.6) & -20.6 (0.1)\\
 \\
 NGC\,1381 &  $82.5^{\phantom{1}82.9}_{\phantom{1}82.2}$ & $0.57^{\phantom{1}0.58}_{\phantom{1}0.55}$ & $ 7.4^{\phantom{1} 7.7}_{\phantom{1} 7.3}$ & $20.8^{\phantom{1}21.6}_{\phantom{1}20.3}$ & $3.1^{\phantom{1}3.2}_{\phantom{1}3.0}$ & 35.2 & 64.2 \{8.7\} & 278.8 (32.4) & -19.0 (0.1)\\
 \\
 IC\,1963  &  $84.6^{\phantom{1}84.7}_{\phantom{1}84.3}$ & $0.39^{\phantom{1}0.42}_{\phantom{1}0.36}$ & $22.6^{\phantom{1}24.9}_{\phantom{1}21.2}$ & $15.8^{\phantom{1}16.1}_{\phantom{1}15.5}$ & $3.6^{\phantom{1}3.7}_{\phantom{1}3.5}$ & 5.5 & 47.8 \{2.1\} & 164.7 (16.8) & -18.5 (0.1)\\
 \\
 NGC\,1375 &  $67.5^{\phantom{1}68.3}_{\phantom{1}66.3}$ & $0.19^{\phantom{1}0.21}_{\phantom{1}0.17}$ & $ 3.2^{\phantom{1} 3.4}_{\phantom{1} 2.9}$ & $14.8^{\phantom{1}15.2}_{\phantom{1}14.0}$ & $2.2^{\phantom{1}2.3}_{\phantom{1}2.1}$ & 10.7 & 35.9 \{11.2\} & 113.4 (19.6) & -18.2 (0.2)\\
 \\
 NGC\,1380A&  $78.1^{\phantom{1}78.8}_{\phantom{1}77.4}$ & $0.18^{\phantom{1}0.24}_{\phantom{1}0.14}$ & $10.9^{\phantom{1}14.5}_{\phantom{1} 8.1}$ & $19.8^{\phantom{1}20.4}_{\phantom{1}19.2}$ & $3.7^{\phantom{1}3.8}_{\phantom{1}3.6}$ & 5.5 & 47.9 \{4.4\} & 119.7 (3.0) & -18.1 (0.2)\\
 \\
 ESO\,358$-$G006 & $65.9^{\phantom{1}67.2}_{\phantom{1}64.8}$ & $0.02^{\phantom{1}0.04}_{\phantom{1}0.01}$ & $ 1.5^{\phantom{1} 2.9}_{\phantom{1} 0.2}$ & $ 9.2^{\phantom{1} 9.6}_{\phantom{1} 8.8}$ & $3.0^{\phantom{1}3.3}_{\phantom{1}2.7}$ & 0.00627 & 29.4 \{19.6\} & 155.8: (26.0) & -17.5 (0.2)\\
 \\
 ESO\,358$-$G059 & $61.9^{\phantom{1}63.9}_{\phantom{1}59.8}$ & $0.72^{\phantom{1}0.76}_{\phantom{1}0.65}$ & $16.2^{\phantom{1}18.4}_{\phantom{1}14.7}$ & $ 2.3^{\phantom{1} 2.5}_{\phantom{1} 2.2}$ & $2.0^{\phantom{1}2.2}_{\phantom{1}1.6}$  & -- & 14.2 \{0.9\} & 108.5: (53.8)  &-17.4 (0.2)\\
 \\
 ESO\,359$-$G002 & $49.4^{\phantom{1}51.7}_{\phantom{1}47.2}$ & $0.070^{\phantom{1}0.14}_{\phantom{1}0.029}$ & $36.0^{\phantom{1}41.7}_{\phantom{1}24.8}$ & $7.8^{\phantom{1}8.2}_{\phantom{1}7.4}$ & $3.1^{\phantom{1}3.4}_{\phantom{1}2.9}$ & 0.4 & 13.9 \{0.4\} & -- &-17.3 (0.2)\\
 \hline
  \end{tabular}\\
 \end{center}
 \footnotesize{Note: For all pertinent calculations,  $H_0=70\,$km$\,$s$^{-1}\,$Mpc$^{-1}$. From (2) to (6), $99\%$ confidence  intervals are presented; from (9) to (10), $1\,\sigma$ errors between  ``()''. Col (2), inclination angle with respect to the line-of-sight; col  (3), bulge to total fraction; col (4), effective radius; col (5),  exponential disk scale length; col (6), S\'ersic index; col (7), minimum  radius where the ``projected velocities correction'' is applied; col (8),  distance of the outermost bin in arc-seconds and in units of the effective  radius; col (9), maximum circular velocity, values followed by ':' should be  taken as an upper limit only; col (10), absolute magnitude in B-band,  assuming a redshift of 0.0043 for the cluster, according to Madore et al. (1999).}
\end{table*}

To convert line-of-sight velocities into rotational motions, we follow
the approach developed by Neistein et al. (1999, hereafter N99).
First of all, allowance has to be made for the extra degree of
``roundness'' contributed by the fact that the disks of S0s are not
infinitely thin; as in N99, we assume that all galaxies would have an
edge-on axis ratio of 0.22, although this assumption turns out not to
be critical to the results.  

The second major effect that must be corrected for is that arising
from projection along the line-of-sight.  Although stars close to the
minimum radius along a line through a highly-inclined disk will have
most of their rotational velocity projected along the line of sight,
those further from the centre of the galaxy will have most of their
motion transverse to the line of sight, so their measured component of
velocity will be small.  This tail of low-velocity stars means that
the measured line-of-sight mean velocity will lie systematically below
the rotational velocity of the stars.  We have therefore applied a correction
for this effect, which is described in Merrett et al. (2006).  This technique can only be
reliably applied in regions where the disk dominates the light, so we
have used the photometric parameters derived above to ascertain the
minimum radius, $R_{\rm LIM}$, at which this correction can be made;
the values for $R_{\rm LIM}$ are given in Table~\ref{tab:photpar}.
Clearly, if $R_{\rm LIM}$ can't be estimated within the radius to which we
have obtained data, then we will not be able to apply this correction, so
will be unable to obtain a reliable rotation curve.  Where it can be
applied, this correction amounts to $\sim 20$\%; although it is
clearly significant, it is still sufficiently small that its exact
value is not critical.

The last correction that we must apply is the difference between mean
streaming velocity, $V_\phi$, and circular rotation speed that arises
because the stars have a random component to their motions as well as
a rotational motion.  This ``asymmetric drift'' can be corrected using
the appropriate Jeans equation (Binney \& Tremaine 1987), and here we follow
the N99 implementation of this correction, by which
\begin{equation}
V_{\rm c}^{2}=V_{\phi}^{2}+\sigma_{\rm fit}^{2}\left(2\frac{R}{R_{\rm
exp}}-1\right),\label{eq:ADC}
\end{equation}
where $\sigma_{\rm fit}$ is the velocity dispersion at radius $R$ determined
from a fit to the velocity dispersion profile using a low-order
polynomial (see Appendix~\ref{app:data}). 

Strictly speaking, this low-order correction is only valid for rather
cold disks with $V_\phi \gg \sigma_{\rm fit}$.  Some experimenting
with simulated data revealed that the corrections remained fairly
reliable for any galaxy with $V_\phi/\sigma_{\rm fit} > 2$, but that
artifacts from the correction would begin to appear in the rotation
curve below this limit.  We can meet this requirement for five of the
sample galaxies, but for two (ESO~356-G006 and ESO~358-G059), we must
relax the limit to $V_\phi/\sigma_{\rm fit} > 1$.  In these cases, the
asymmetric drift correction of equation~\ref{eq:ADC} tends to be too
large, so the resulting rotation velocities should strictly be taken
as only an upper limit.  In two cases (NGC~1316 and ESO~359-G002),
random motions dominate at all radii, so this correction cannot even
be attempted.  In consequence, these two galaxies have to be excluded from
subsequent analysis.

\begin{figure*}
  \includegraphics[scale=0.47]{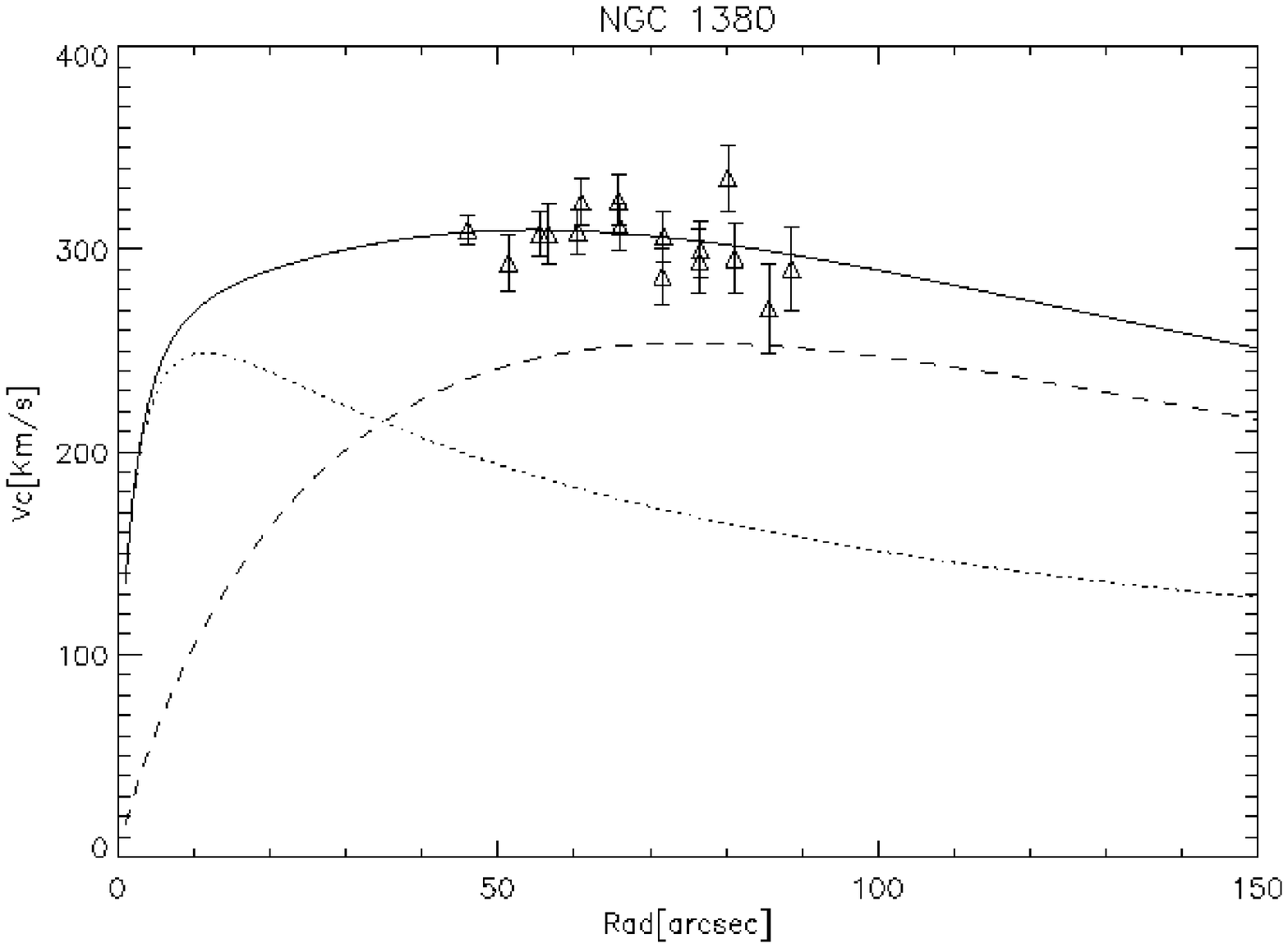}
  \includegraphics[scale=0.47]{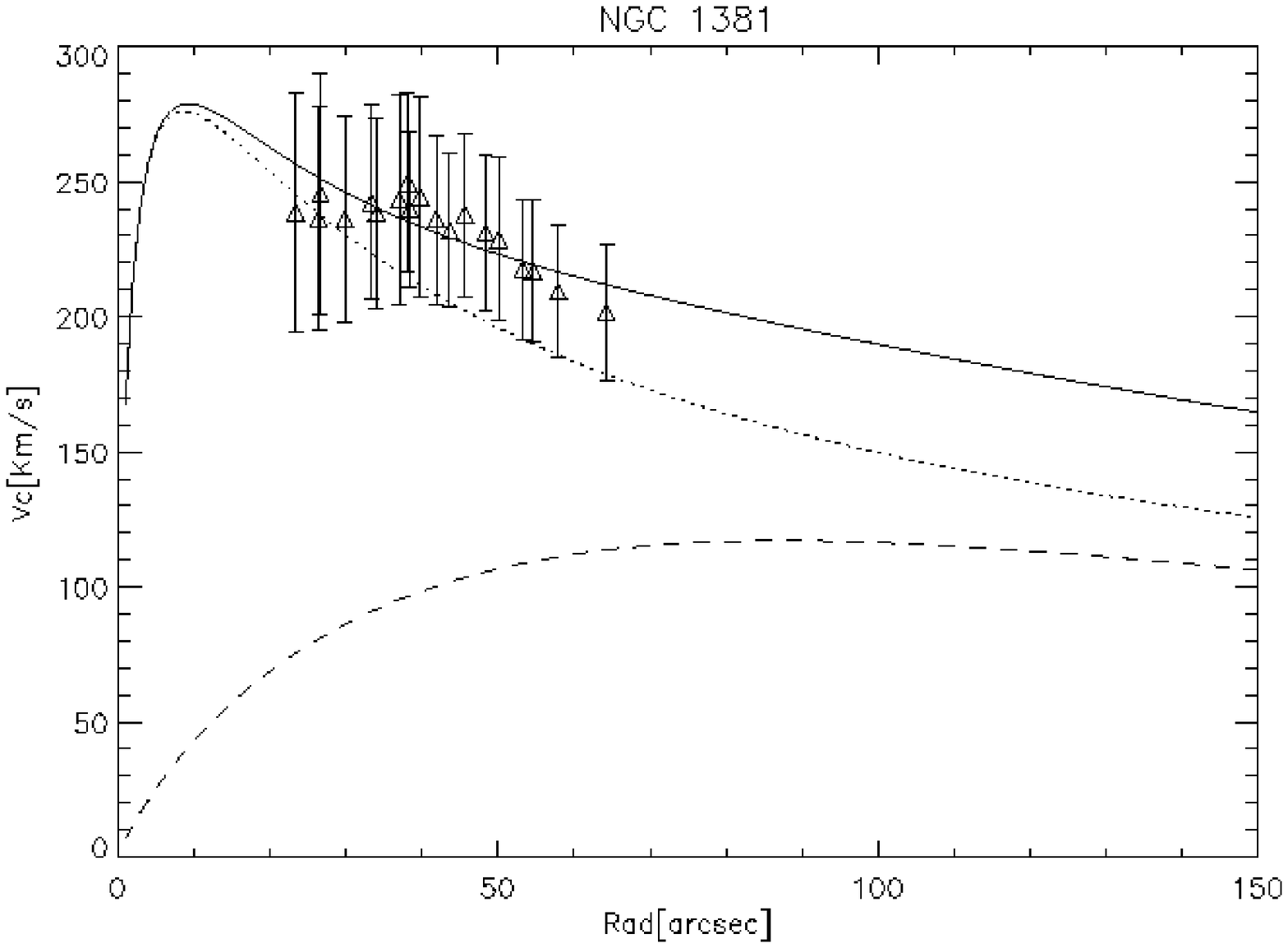}
  \includegraphics[scale=0.47]{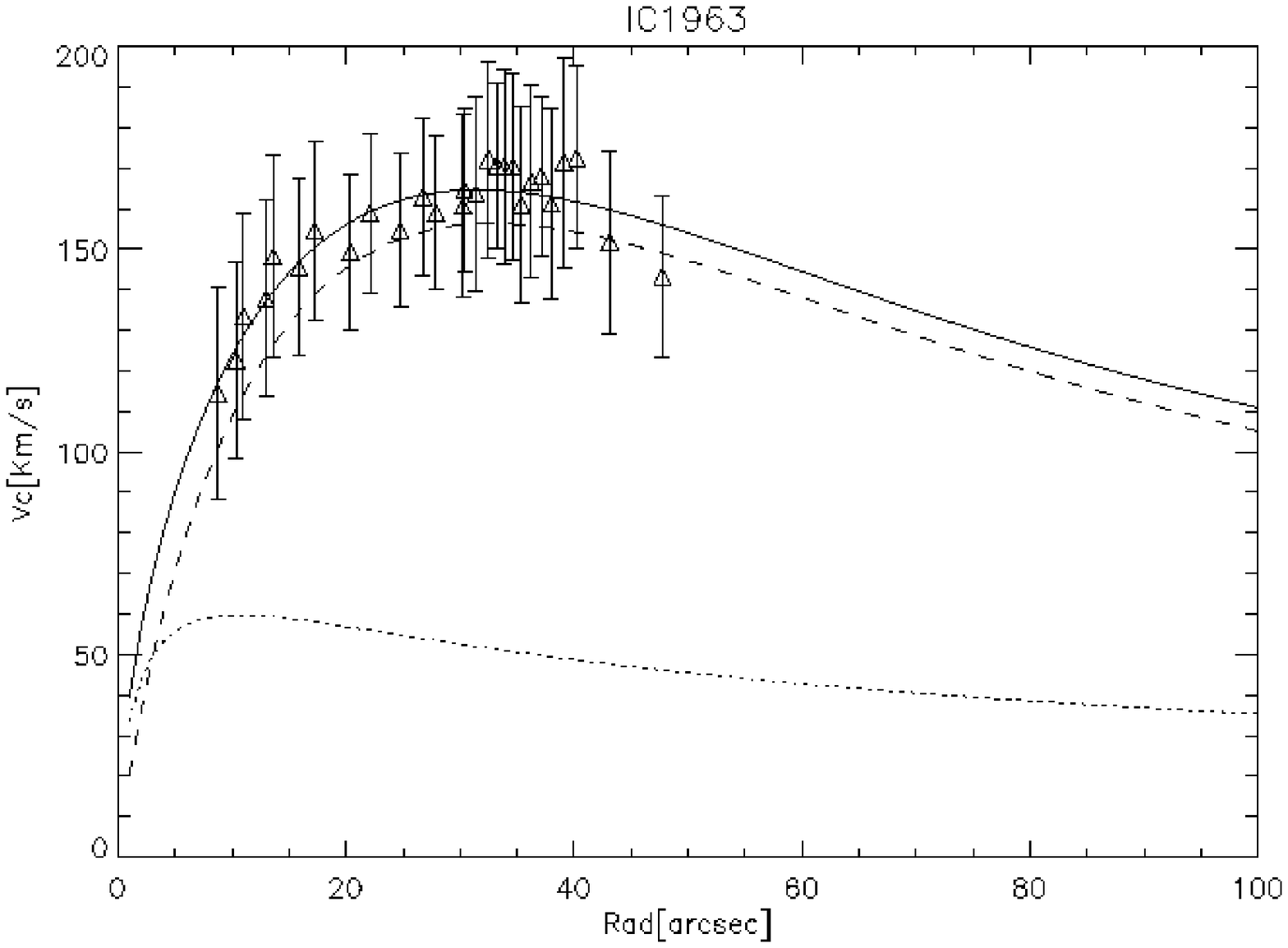}
  \includegraphics[scale=0.47]{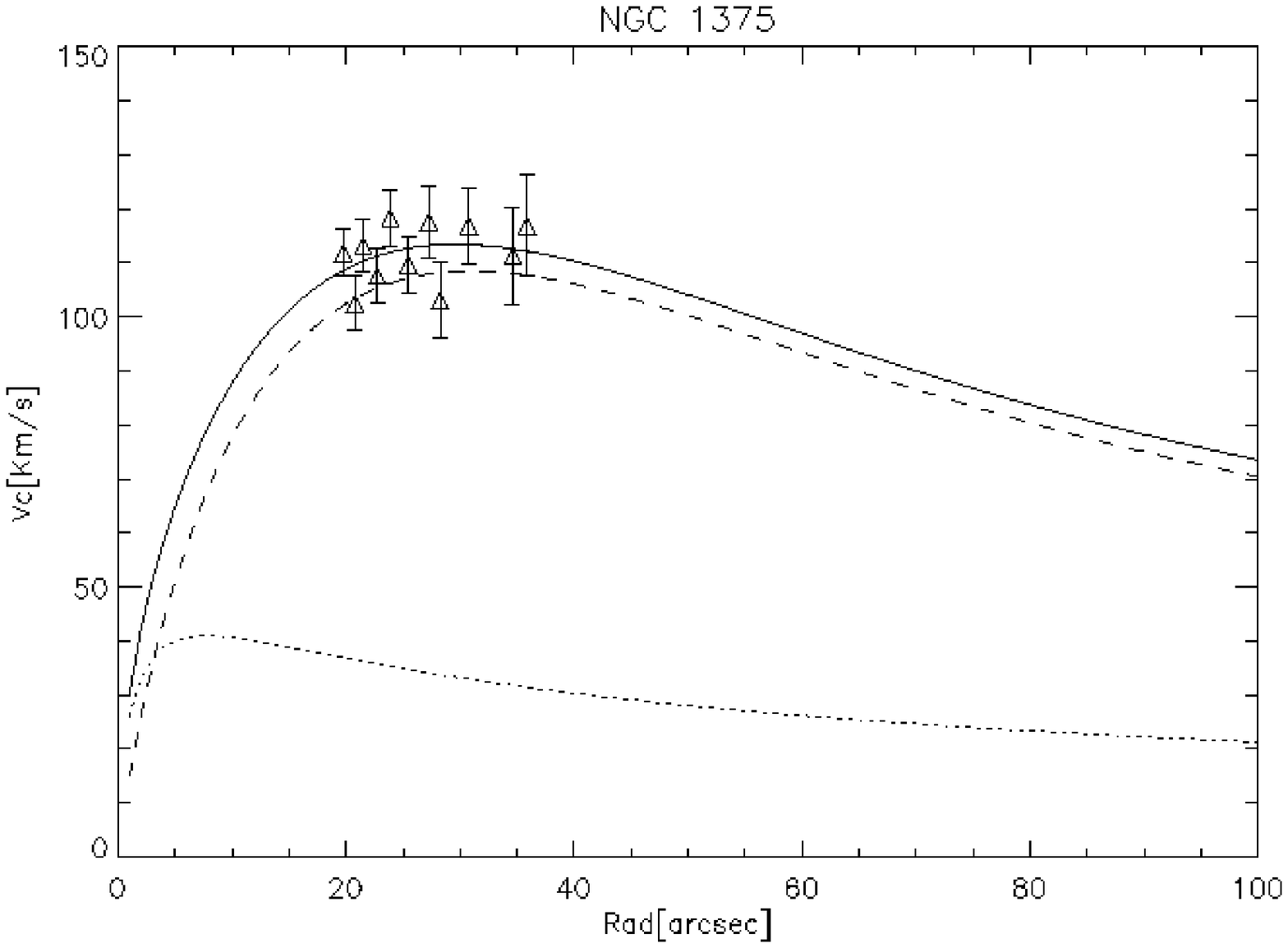}
  \includegraphics[scale=0.47]{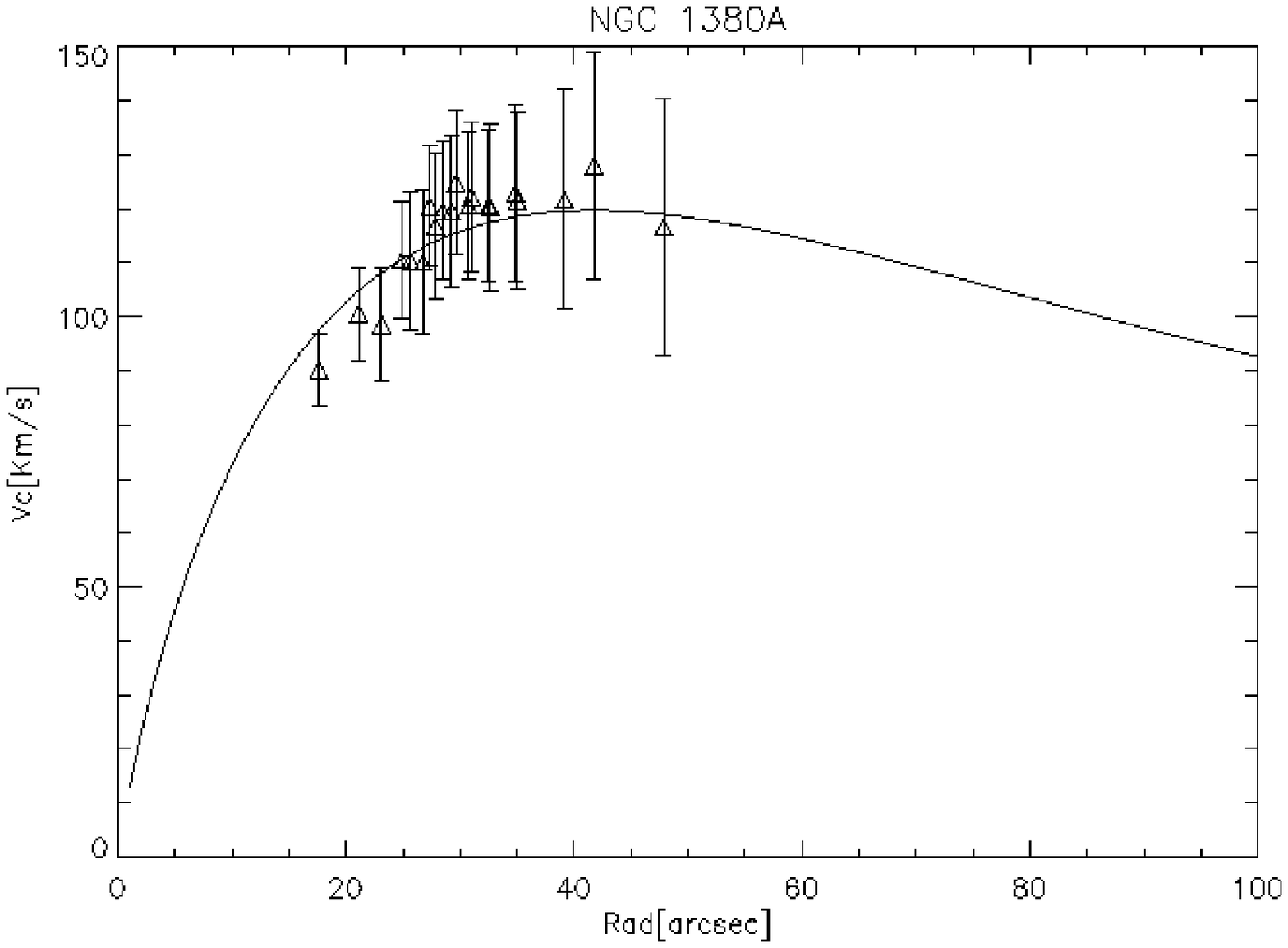}
  \caption{Derived circular velocity as function of radius and the
  corresponding best-fit models for Fornax Cluster S0 galaxies.  The
  dotted line shows the bulge contribution, the dashed line shows the
  disk contribution, and the solid line shows the quadrature-summed
  total. For NGC~1380A a pure disk model gives the best fit.}\label{fig:vcfit}
\end{figure*}
After carrying out the above procedures, we can convert the measured
mean line-of-sight velocity data points into discrete estimates of the
local circular speed at the corresponding radii.  These estimates have
two limitations: first, each point will have a significant
error bar from all the propagated uncertainties; and second, the data
will lie over only a limited range in radii since the rotation
velocities cannot be determined in this way at small radii where
random motions dominate, and the galaxies become too faint at large
radii to have their kinematics measured.  We therefore cannot simply
read off dynamical parameters such as the maximum velocity, $V_{\rm
max}$, from these data.  Instead, we effectively interpolate and
extrapolate the limited data by modelling the contributions to the
rotation curve by disk and bulge.  Using the photometric parameters
derived above, we model the disk component of the rotation curve using
the formula for a pure exponential,
\begin{equation}
   V_{c}^{2}(R)=4\pi G \Sigma_0 R_{exp}y^2[I_0(y)K_0(y)-I_1(y)K_1(y)],
\end{equation}
where G is the gravitational constant, $\Sigma_0$ is the central
surface density, $I_0, K_0, I_1$ and $K_1$ are modified Bessel
functions, and $y=R/R_{\rm exp}$ (Freeman 1970).  For the bulge, we
use the rotation curve of Hernquist (1990),
\begin{equation}
   V_{c}^{2}(R)=\frac{G M_{bulge}}{R}\frac{R^2}{(R+a)^2},
\end{equation}
where $M_{\rm bulge}$ is the mass of this component, and the
scalelength $a$ can be approximately related to $R_{\rm e}$ by
$a/R_{\rm e} = 1.82$. The normalizations of these two components
(effectively, the mass-to-light ratios of the photometric components)
were then varied such that their sum produced the best fit to the
circular speed data over the range that it is available.  The result
of this process is presented in Figure~\ref{fig:vcfit}; we have not
attempted this procedure for the two galaxies in which the circular
velocities are very uncertain because the asymmetric drift corrections
were too large.

We can now use these fits to read off quantities of interest.  For
example, for a Tully--Fisher analysis, we need the maximum rotation
velocity $V_{\rm max}$.  The resulting values are presented in
Table~\ref{tab:photpar}; for the two galaxies for which we were unable
to carry out the complete fit, we simply quote the maximum circular
speed as derived from the data.  Note that although we have generally
derived $V_{\rm max}$ from the fitting procedure, it is apparent from
Figure~\ref{fig:vcfit} that the values returned do not differ greatly
from the measured rotation speeds.  This similarity arises because the
new VLT observations go deep enough to enable us, for the first time,
to measure velocities in the flat part of the rotation curve.  By
reaching these radii, we enter a regime in which the maximum velocity
is reasonably tightly constrained irrespective of the details of the
model fit.

\section{Summary}\label{sec:conc}

In this paper, we have presented kinematics derived from VLT long-slit
spectral observations of 9 S0 galaxies in the Fornax Cluster.  The
derived kinematic parameters are the mean line-of-sight velocity,
velocity dispersion, and the higher-moment $h_3$ and $h_4$
coefficients.  The deep nature of the spectra mean that we have
obtained these parameters out to distances equal or larger than 2$\, R_e$ of
the bulge for the majority of our objects, so we are typically probing the
disk-dominated part of the kinematics, allowing us to study both disk and
bulge dynamics in some detail.

A first qualitative pass through these kinematic data for the
individual galaxies indicates that some of the existing data, due to
both their lower signal-to-noise ratio and poorer spectral resolution,
are unreliable.  It also reveals that there is quite a lot of
complexity in these systems, suggesting that their evolutionary
histories may be relatively complex.  With the quality of these new
data, we should be able to obtain significant new archaeological clues
as to how they evolved to their current states.  As a first step in
this direction, we have estimated one of the principal defining
dynamical characteristics of these systems, their maximum rotation
speeds.  In the next paper, we will use this quantity to explore the
evolutionary clues in these systems' Tully--Fisher relation, while in
a subsequent paper we will exploit the high quality of the spectra to
investigate the further clues that may be provided by the composition
of their stellar populations out to large radii.

\section*{Acknowledgements}
We would like to thank Dr. Osamu Nakamura, Dr. Steven
Bamford, Dr. Mustapha Mouchine, Dr. Jes\'us Falc\'on-Barroso, Dr. Reynier
Peletier and Dr. Konrad Kuijken for their
help, suggestions and interesting discussions. This work was based on
observations made with ESO telescopes at Paranal Observatory under
programme ID 070.A-0332. The Dark Cosmology Centre is funded by the Danish
National Research.

\appendix

\section{Line-of-sight Kinematics:}\label{app:data}
This appendix shows plots of the major-axis kinematics for the sample
of 9 Fornax S0 galaxies.  The values of mean velocity, $V_{\rm LOS}$,
velocity dispersion, $\sigma$, and higher-moment coefficients, $h_3$
and $h_4$ are plotted as a function of radius, and the location of
each galaxy's effective radius, $R_{\rm e}$, is also marked as a
dashed line.  The values of $R_{\rm e}$ and the position angle of the
slit are annotated on each plot.  The solid lines show the fit to the
dispersion profile, $\sigma_{\rm fit}(R)$, adopted in
Section~\ref{sec:vcirc}

\begin{figure*}
\centering
\centering
\includegraphics[scale=0.5]{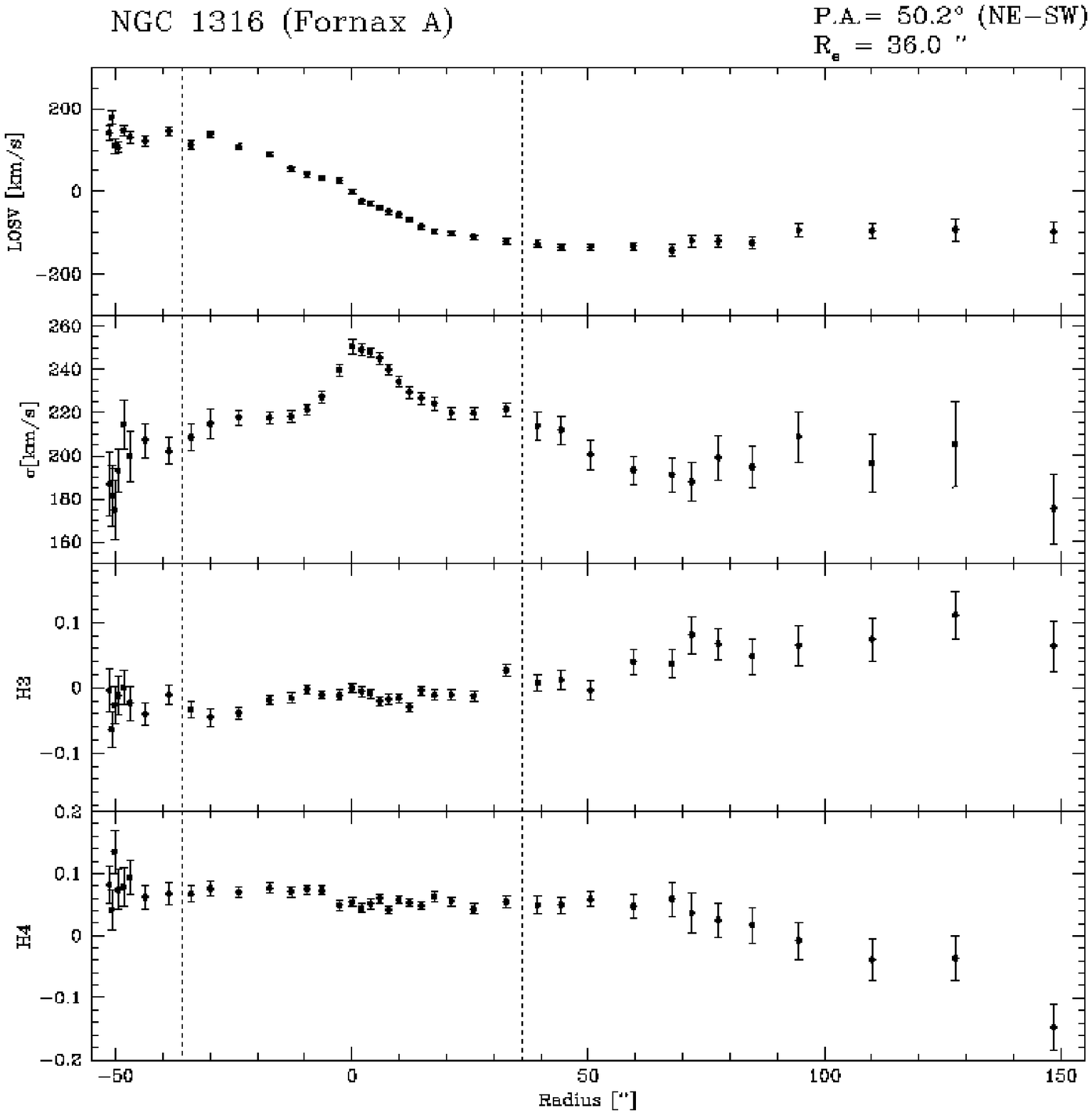}
\caption{The major-axis kinematics of NGC~1316.}\label{fig:1316dat}
\end{figure*}
\begin{figure*}
\centering
\centering
\includegraphics[scale=0.5]{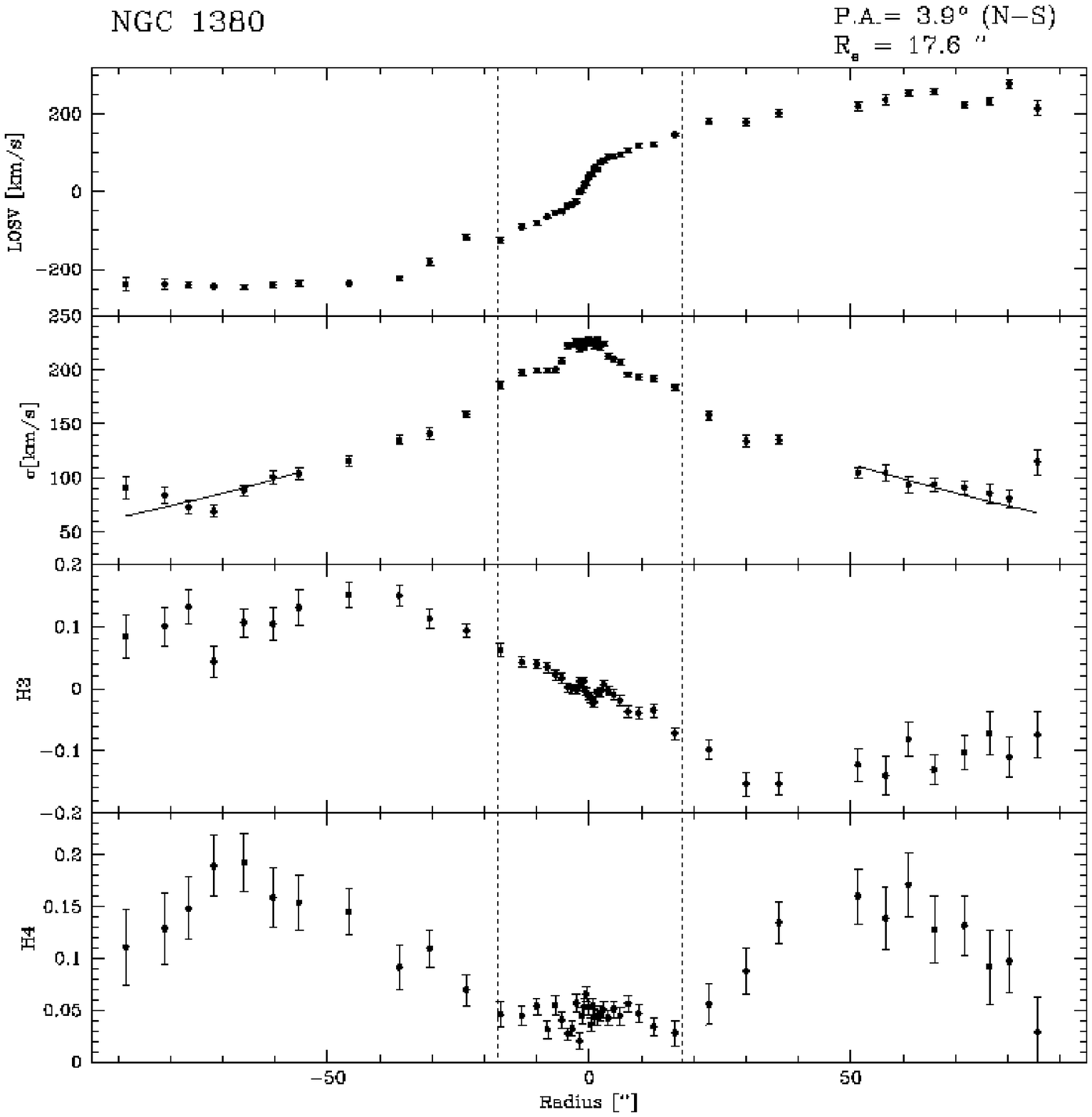}
\caption{The major-axis kinematics of NGC 1380.}\label{fig:1380dat}
\end{figure*}
\begin{figure*}
\centering
\centering
\includegraphics[scale=0.5]{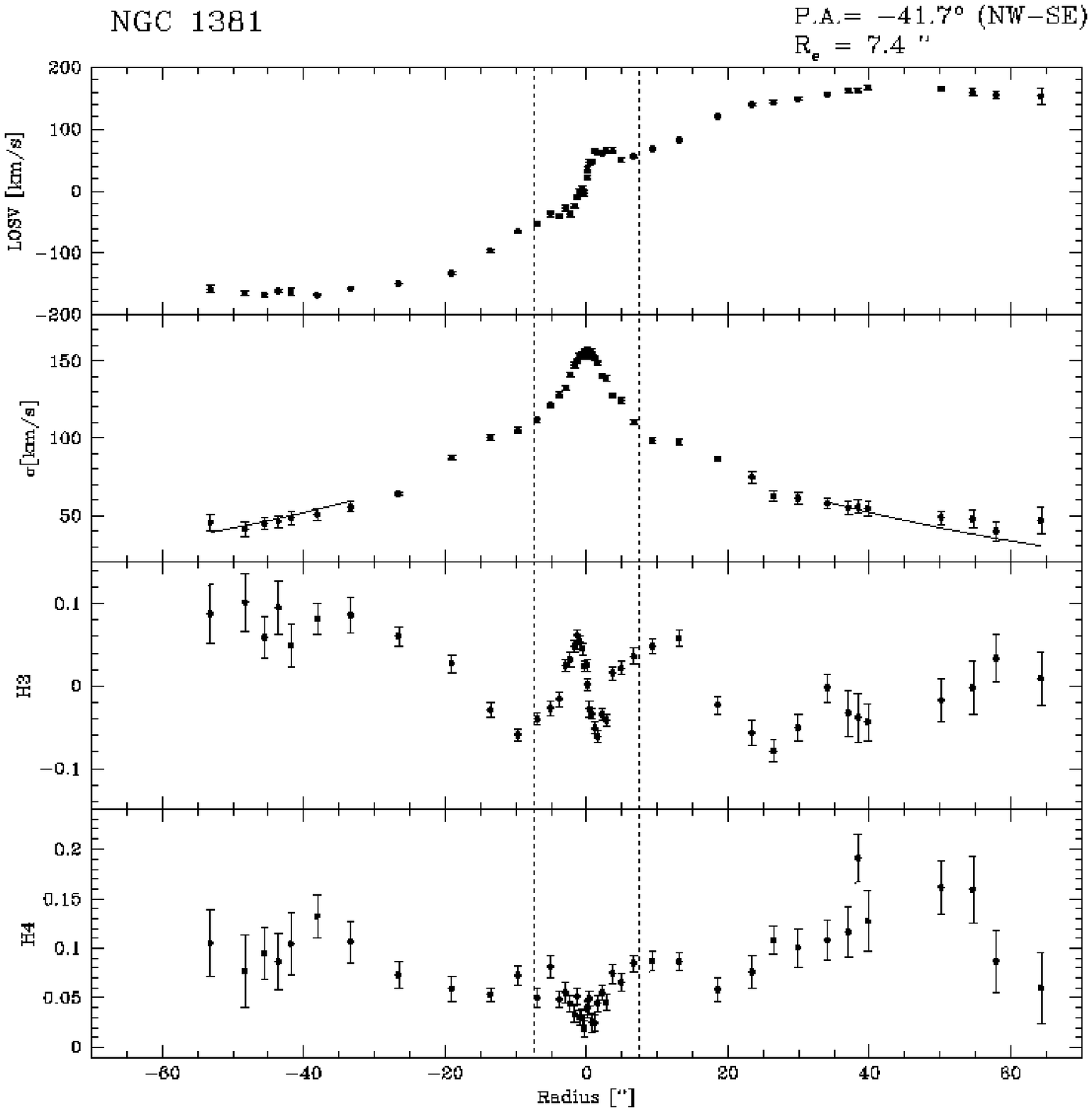}
\caption{The major-axis kinematics of NGC 1381.}\label{fig:1381dat}
\end{figure*}
\begin{figure*}
\centering
\centering
\includegraphics[scale=0.5]{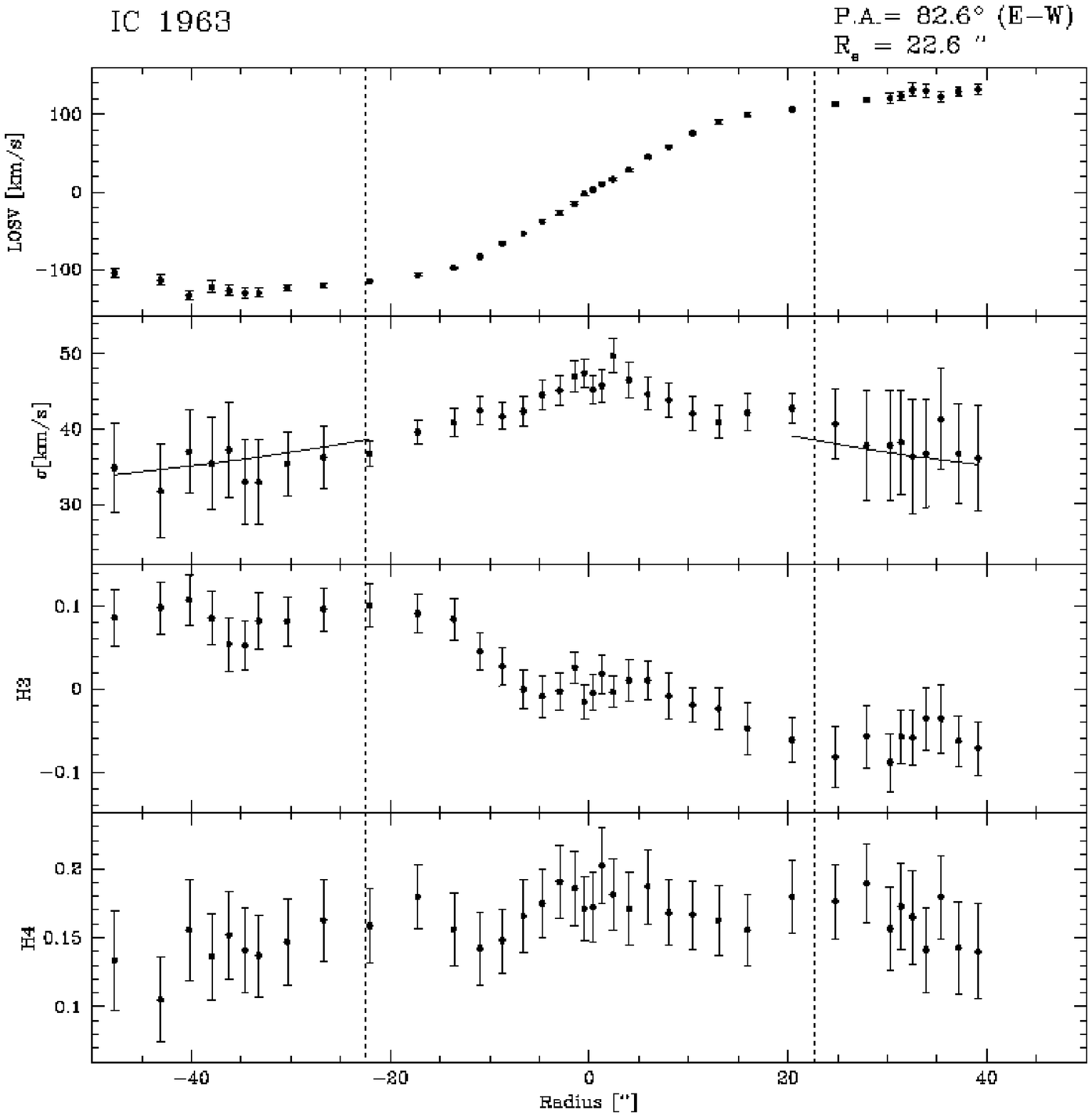}
\caption{The major-axis kinematics of IC 1963.}\label{fig:1963dat}
\end{figure*}
\begin{figure*}
\centering
\centering
\includegraphics[scale=0.5]{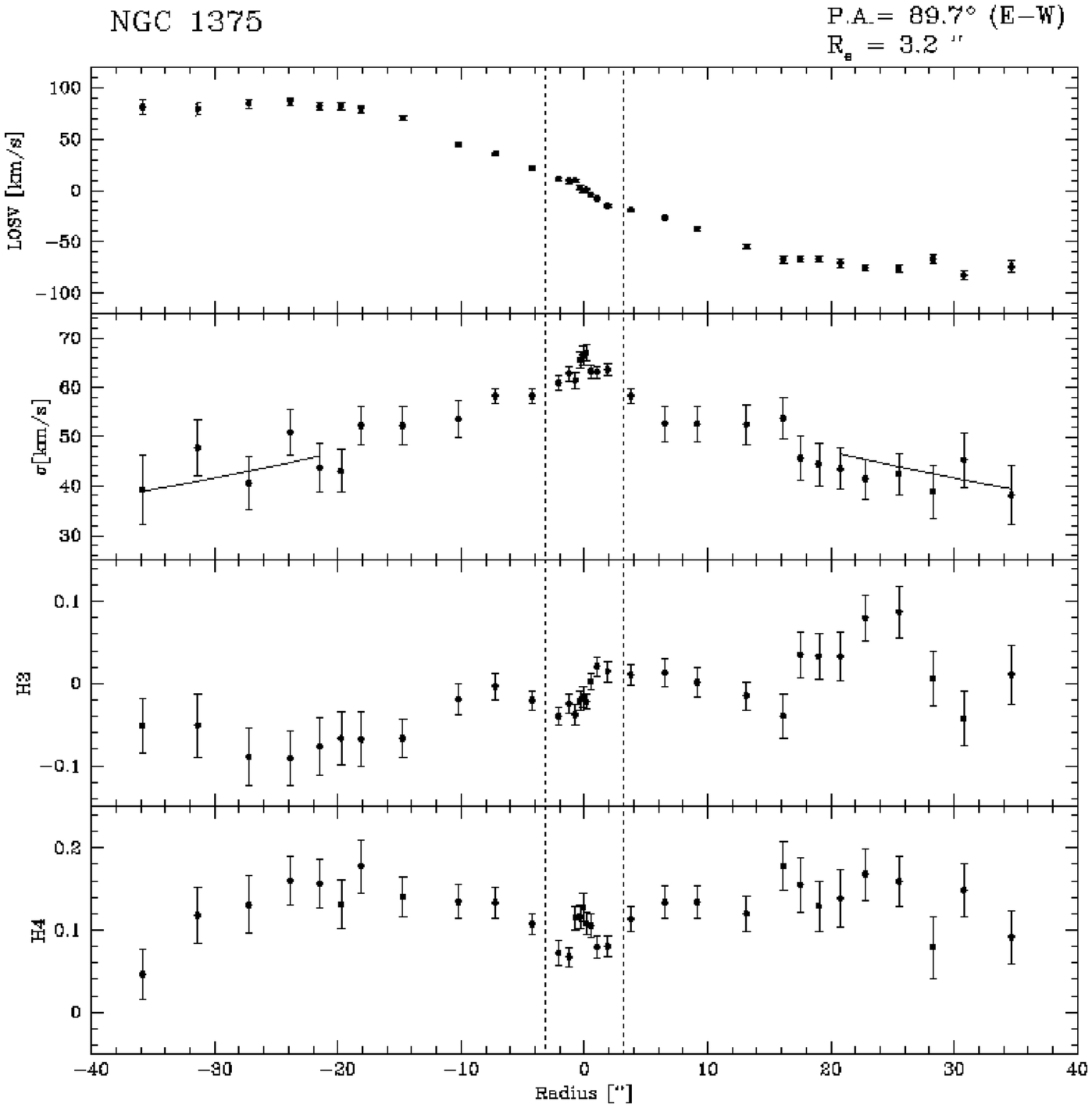}
\caption{The major-axis kinematics of NGC 1375.}\label{fig:1375dat}
\end{figure*}
\begin{figure*}
\centering
\centering
\includegraphics[scale=0.5]{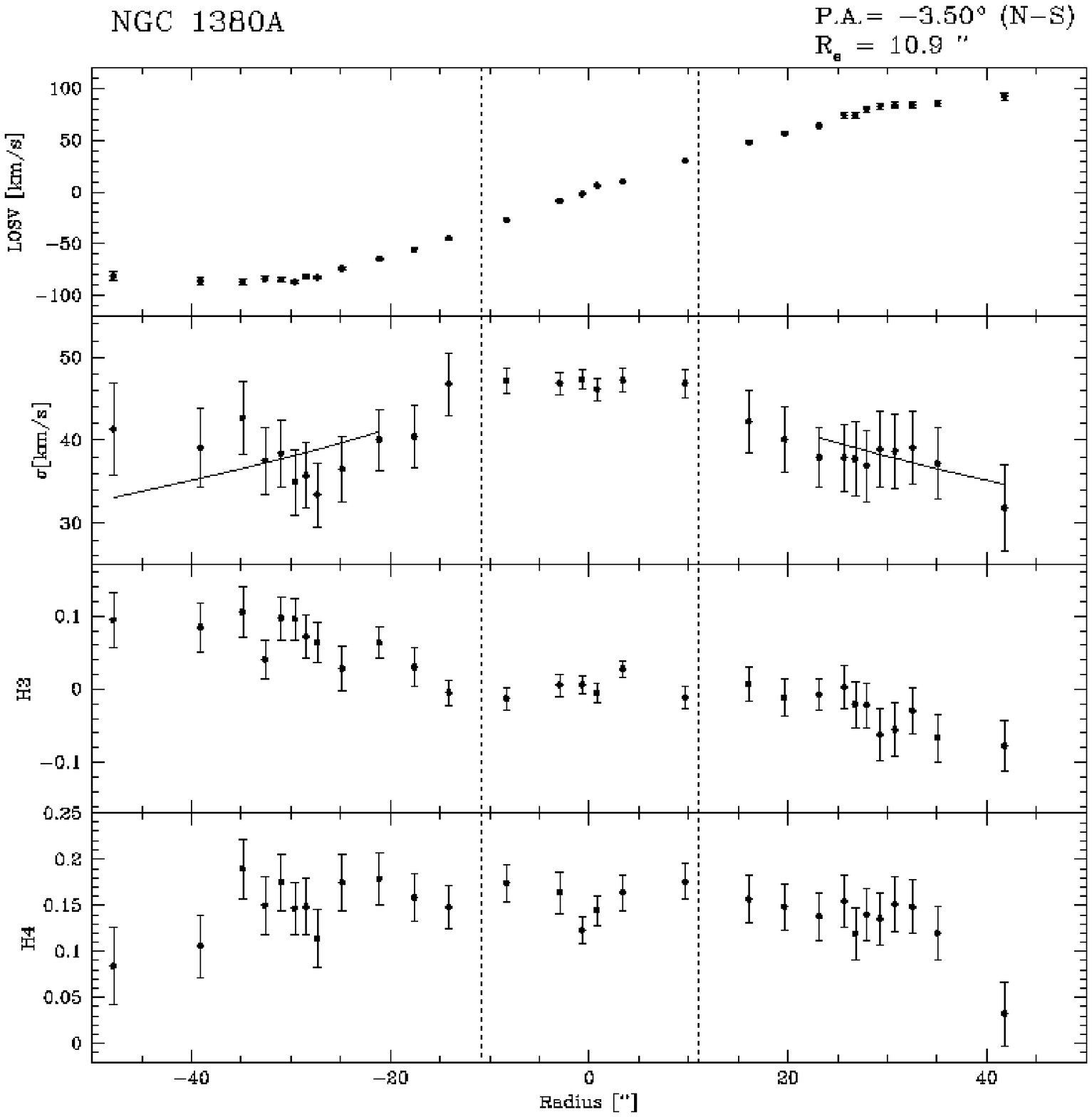}
\caption{The major-axis kinematics of NGC 1380A.}\label{fig:1380Adat}
\end{figure*}
\begin{figure*}
\centering
\centering
\includegraphics[scale=0.5]{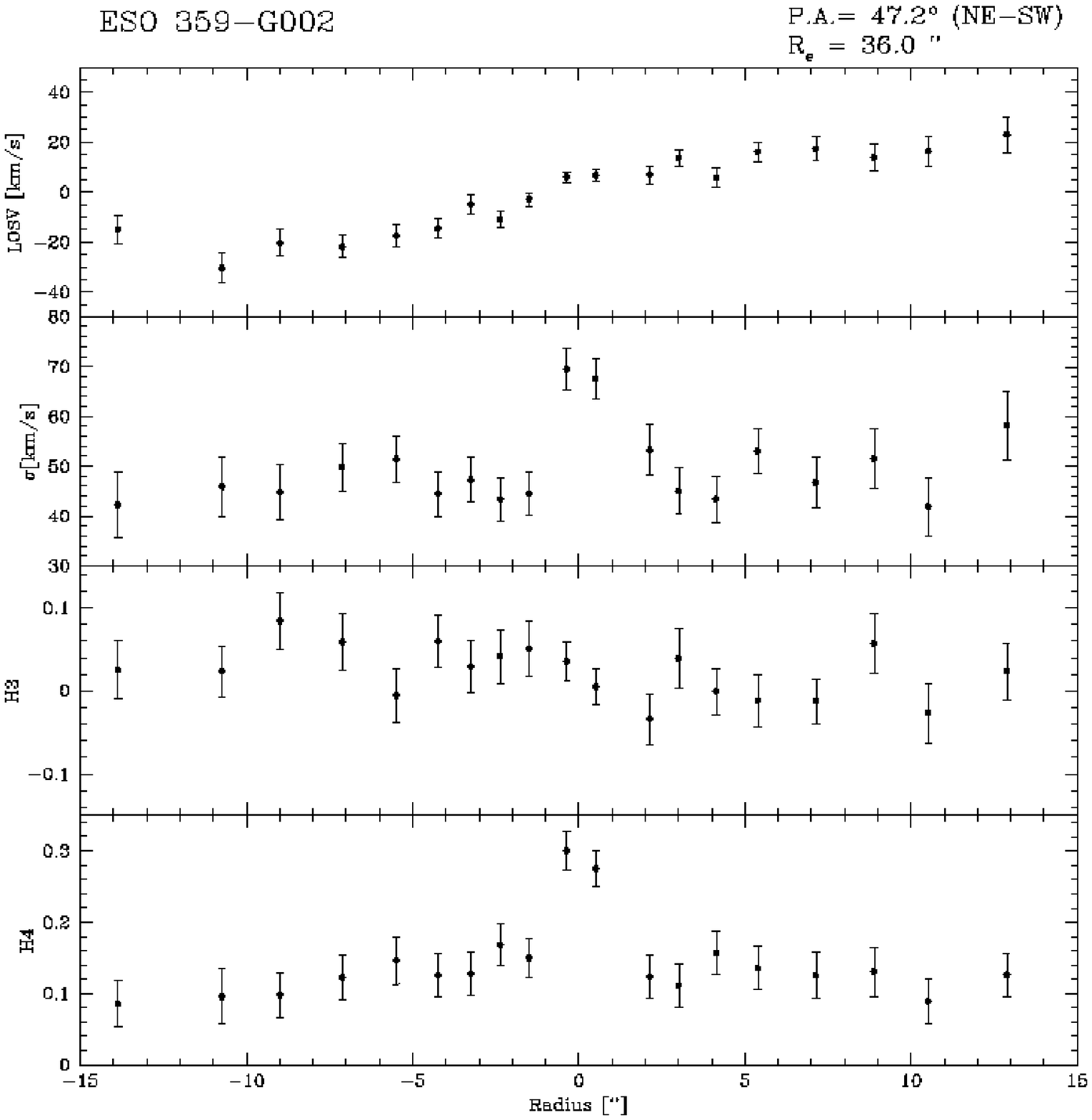}
\caption{The major-axis kinematics of ESO 359-G002.}\label{fig:G002dat}
\end{figure*}
\begin{figure*}
\centering
\centering
\includegraphics[scale=0.5]{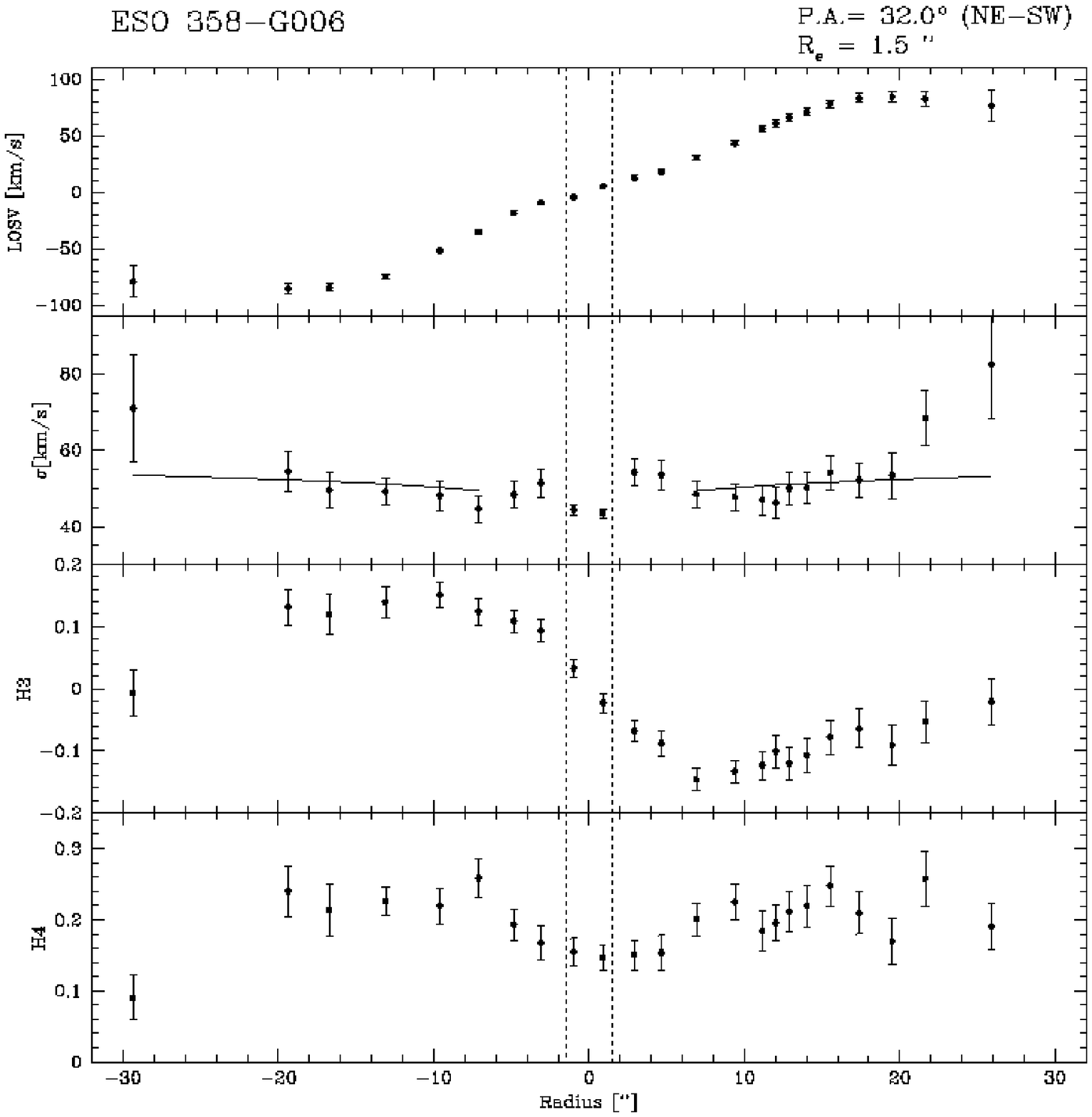}
\caption{The major-axis kinematics of ESO 358-G006.}\label{fig:G006dat}
\end{figure*}
\begin{figure*}
\centering
\centering
\includegraphics[scale=0.5]{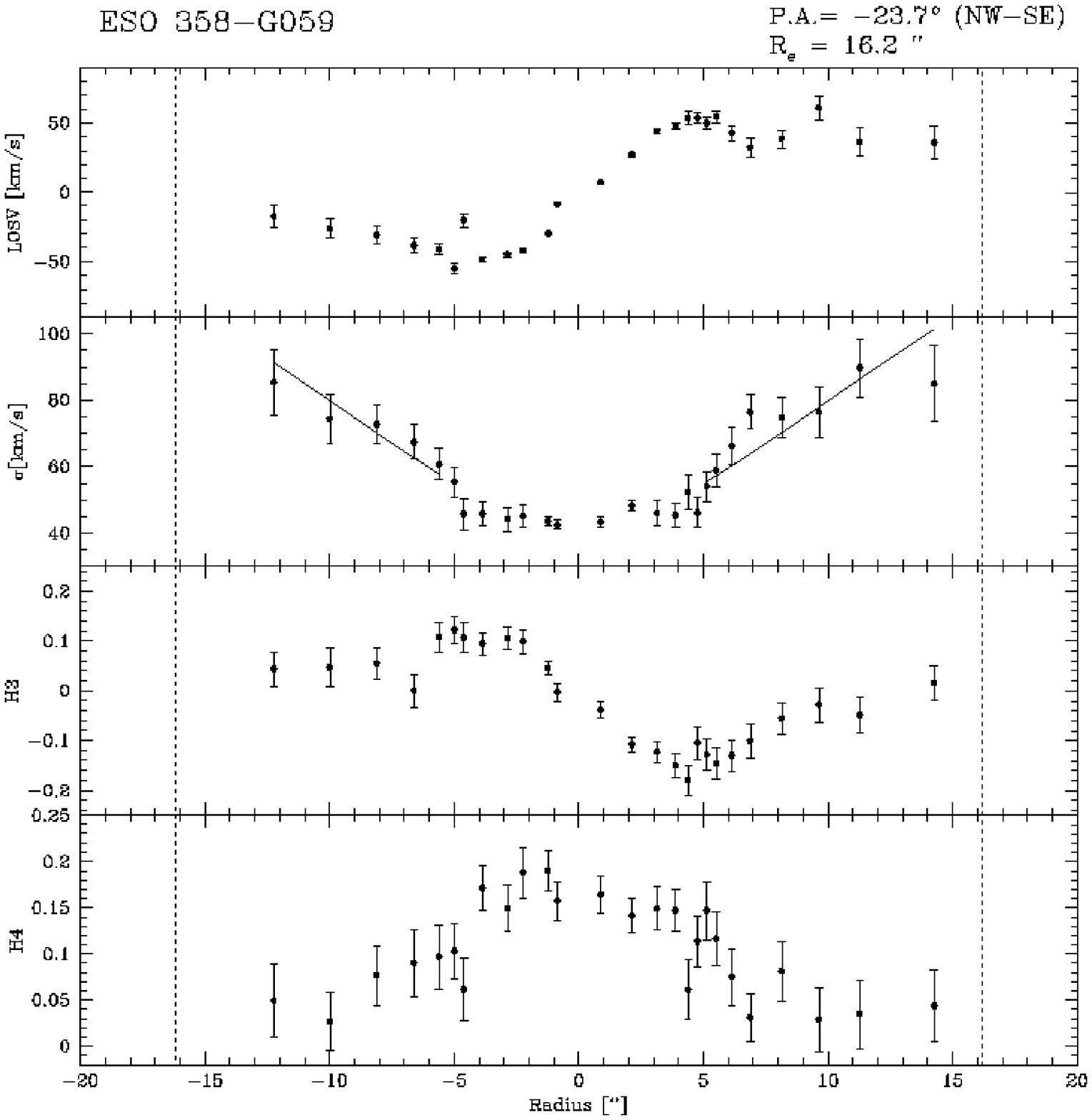}
\caption{The major-axis kinematics of ESO 358-G059.}\label{fig:G059dat}
\end{figure*}

\label{lastpage}

\end{document}